\documentclass[10pt]{iopart}
\usepackage{epsf}
\usepackage{appendix}

\newcommand{\bra}[1]{\left(#1\right)}
\newcommand{\bras}[1]{\left[#1\right]}

\newcommand{\hatcal}[1]{\hat{\cal{#1}}}
\newcommand{\sla}[1]{\displaystyle{\not}#1}
\newcommand{\slacal}[1]{{\sla{\cal{#1}}}}
\newcommand{\be}{\begin{equation}}
\newcommand{\ee}{\end{equation}}
\newcommand{\ber}{\begin{eqnarray}}
\newcommand{\eer}{\end{eqnarray}}

\begin{document}

\title{Density growth in Kantowski-Sachs cosmologies with cosmological constant}
\author{Michael Bradley$^1$, Peter K.S. Dunsby$^2$, Mats Forsberg$^1$ and Zolt\'an Keresztes$^3$}
\address{$^1$ Department of Physics, Ume\aa\ University, Sweden}
\address{$^2$ Department of Mathematics and Applied Mathematics, University of Cape Town, South Africa}
\address{$^3$ Department of Experimental Physics, University of Szeged, Hungary}
\ead{michael.bradley@physics.umu.se, peter.dunsby@uct.ac.za, mats.forsberg@physics.umu.se,
zkeresztes@titan.physx.u-szeged.hu}

\date{\today }

\begin{abstract}
In this work the growth of density perturbations in Kantowski-Sachs cosmologies with a positive 
cosmological constant is studied, using the 1+3 and 1+1+2 covariant formalisms. For each wave 
number we obtain a closed system for scalars formed from quantities that are zero on the background 
and hence are gauge-invariant. The solutions to this system are then analyzed both analytically and
numerically. In particular the effects of anisotropy and the behaviour close to a bounce in the cosmic 
scale factor are considered.
We find that typically the density gradient in the bouncing direction experiences a local maximum at 
or slightly after the bounce.
\end{abstract}

\submitto{CQG}
\maketitle
%%%%%%%%%%%%%%%%%%%%%%%%%%%%%%%%%%%%%%%%%%%%%%%%
\section{Introduction}
%%%%%%%%%%%%%%%%%%%%%%%%%%%%%%%%%%%%%%%%%%%%%%%%
The observed distribution of inhomogenities and anisotropies in the background radiation seems to be 
well described by the $\Lambda$CDM model, see e.g. \cite{Spergel48, Komatsu49}.  However they do not match  
perfectly with this model, \cite{Bennett46, Oliveria45, Vielva59},  and consequently it is of interest to explore the 
complete phase-space of cosmological models and to see how differences from the standard picture affect the basic properties 
of the universe. 

Perturbations of  anisotropic cosmological models have been considered by many authors, e.g., 
\cite{Doroschkevich, Perko, Hu, Abbott, Tomita}, using methods depending on the choice of gauge,
like the perturbation theory of Lifshitz and Khalatnikov \cite{Lifshitz} or Bardeen's gauge invariant 
theory \cite{Bardeen}. 
Unfortunately, the variables in  Bardeen's theory are defined with respect to a particular coordinate system, 
making their geometrical and physical meaning unclear \cite{Stewart}. In the covariant approaches one circumvents 
these problems by using the spatial curvature rather than the metric as defining variables \cite{Hawking, Olson} 
and as the perturbed variables choosing objects that are zero on the background \cite{EllisBruni, BruniDunsbyEllis}, 
and hence are gauge invariant \cite{StewartWalker}.
Some works on perturbations in anisotropic cosmological models along these lines can be found in
\cite{Dunsby, Osano}.

Bouncing cosmologies are of interest since the initial
singularity can be avoided in these models. The observable universe seems to be close to
homogeneous and isotropic, see e.g. \cite{Komatsu}. Hence it may be approximated by the 
Friedmann-Lema\^{\i}tre-Robertson-Walker (FLRW) metric that also serves as
isotropic limits of different Bianchi models. In particular, the flat, open and
closed FLRW universes are isotropic limits of Bianchi I, V and IX models,
respectively \cite{Maccallum}. From the Raychaudhuri equation it follows that a bounce is not
possible in FLRW models if the strong energy condition is required to hold \cite{Ellis}.
Nevertheless, when the weak energy condition holds a bounce can occur in a closed
FLRW universe \cite{GoliathEllis}, \cite{closedUniv}, while matter violating the
null energy condition can cause a bounce in flat or open universes \cite%
{exotic1}-\cite{exotic4}. Bouncing FLRW universes are also suggested by
alternative theories of gravitation such as $f\left( R\right) $ gravity \cite%
{fR}, $\ f\left( T\right) $ gravity \cite{fT} and braneworlds \cite{brane},
or by loop quantum gravity \cite{loop1}-\cite{bouncingModels}. 

Cosmic
microwave background (CMB) observations support a close to flat FLRW ($\left\vert
\Omega _{k}\right\vert <10^{-2}$) universe \cite{Komatsu}.
Such constraints from the CMB observations have also been considered in bouncing cosmologies.
Some models wherein a
bounce appears due to a scalar field suffer from high tensor-to-scalar ratio 
\cite{exotic2}, \cite{excludeBounce}. However, this can be remedied both in
general relativity \cite{supportBounce} and $f\left( T\right) $ gravity \cite%
{fT} by introducing another massless scalar field. It was shown in loop
quantum cosmology for a bounce followed by an inflationary phase that the
tensor-to-scalar ratio places within the observational bound and that the low CMB
multipoles are suppressed \cite{loopCMB1}-\cite{loopCMB3}. Actually, this
suppression is observed but it can also have other origins \cite%
{PrimordialCMB}.
Bouncing FLRW models with dust, radiation and cosmological constant are excluded
by observations due to an elegant argument by B\"orner and Ehlers \cite{Ehlers}. 
If the bounce took place before the formation of the quasars, that are observed at
larger redshifts than 4, the present fraction of matter relative to the critical one would
be less than 0.02, in contradiction with current estimates.

It is not only the FLRW cosmologies that are interesting as possible models of
the universe. The other models that have been investigated in some depth are those that have  
homogeneous but anisotropic geometries, namely the Bianchi and Kantowski-Sachs 
\cite{KantowskiSachs} cosmologies. Bianchi models were also used successfully to explain
the CMB spectrum \cite{BianchiCMB1}-\cite{BianchiCMB4} and  Kantowski-Sachs
cosmologies might also be relevant for this purpose. The
particular model presented in \cite{KSCMB} can explain some features of the CMB
spectrum, however not all.

The cosmic no-hair conjecture says that spacetimes containing a positive
cosmological constant finally evolves into de Sitter state. Wald proved that this
happens for all Bianchi models with exception of Bianchi IX \cite{Wald}. It
was assumed in the proof that the fluid congruences are orthogonal to the homogeneous
symmetry surfaces and that the matter energy-momentum tensor, not including the
cosmological constant, satisfies the dominant and strong energy conditions.
The isotropisation also happens under inflation provided via a scalar field
with an exponential potential \cite{Kitada}-\cite{Hoogen}. However, in the
tilted case isotropisation does not necessarily take place even during
inflation \cite{GoliathEllis}, \cite{Goliath}. 
For Bianchi IX universes Wald gave a sufficient condition for them to
evolve into de Sitter state. The evolution of Kantowski-Sachs universes with positive
cosmological constant show similarities with that of Bianchi IX cosmologies \cite{Moniz}.
Although not the all initial conditions lead to de Sitter state, Moniz
showed that the cosmic no-hair conjecture is widely valid in Kantowski-Sachs
universes. 
Basically, the Kantowski-Sachs cosmologies with positive
cosmological constant can evolve into de Sitter or Kasner states \cite{GoliathEllis}.
%
%Kantowski-Sachs cosmologies with
%cosmological constant and a perfect fluid having constant barotropic index that
%satisfies the strong energy condition can evolve into de Sitter or Kasner
%states \cite{GoliathEllis}.

In this paper we investigate the density growth in perfect fluid Kantowski-Sachs
cosmologies with positive cosmological constant mainly in those cases when 
isotropisation happens. 
%The background evolution can be various in Kantowski-Sachs cosmologies. 
We consider both bouncing and non-bouncing
types of evolutions. The conditions under which bounces are possible were
studied in \cite{SolomonsDunsbyEllis} and with a positive cosmological constant the
Kantowski-Sachs models may under certain conditions undergo a bounce. This
effect can also be achieved for Kantowski-Sachs solutions in $R^{n}$
gravity, see \cite{LeonSaridakis}.
%In the present paper we study the growth of density perturbations  in perfect fluid Kantowski-Sachs universes with 
%positive cosmological constant with special attention dedicated to the behaviour close to a bounce,
To study density perturbations we use the 1+3 
and 1+1+2 covariant splits of spacetime \cite{EllisBruni,EllisvanElst, ClarksonBarrett, Clarkson}. 
As inhomogeneity variables the spatial gradients of the density, the expansion, the shear scalar and one more 
auxiliary scalar to close the system, are used. These quantities are zero on the background, and are hence gauge invariant. By 
projecting along the preferred and orthogonal directions respectively, taking divergences of these projections
and making harmonic decompositions of the spatial derivatives, the system is reduced to a first order 
system in time of four scalar quantities for
each wave number.

In section \ref{sec:preliminaries} we briefly review the 1+3 and 1+1+2 covariant splits of spacetime. Then in section 
\ref{sec:KantowskiSachs}
the background solutions, of Kantowski-Sachs type, are studied and all background vacuum solutions
are given. The perturbative equations
are determined in section \ref{sec:pert} and some analytical results are obtained in \ref{sec:analytical}.
Numerical studies of perturbations on different backgrounds, both with and without bounces, are performed in section 
\ref{sec:numerical}. In \ref{appE} some results on bouncing closed FLRW universes are given for reference.

%%%%%%%%%%%%%%%%%%%%%%%%%%%%%%%%%%%%%%%%%%%%%%%%
\section{The 1+3 and 1+1+2 covariant formalisms}\label{sec:preliminaries}
%%%%%%%%%%%%%%%%%%%%%%%%%%%%%%%%%%%%%%%%%%%%%%%%

The 1+3 and 1+1+2 covariant splits of spacetime are here briefly reviewed. These formalisms are 
suitable also for perturbative calculations, as will be done in section \ref{sec:pert}. 
For more details see, e.g.,
\cite{EllisBruni, EllisvanElst} and \cite{ClarksonBarrett, Clarkson}.
%%%%%%%%%%%%%%%%%%%%%%%%%%%%%%%%%%%%%%%%%%%%%%%%
\subsection{Preliminaries}
%%%%%%%%%%%%%%%%%%%%%%%%%%%%%%%%%%%%%%%%%%%%%%%%
In \cite{EllisBruni, EllisvanElst} a covariant formalism for the 1+3 split of spacetimes with a preferred time-like
vector, $u^a$, was developed. The projection operator onto the perpendicular 3-space is given by
\begin{equation}
h_a^b=g_a^b+u_au^b \, .
\end{equation}
Projections with $h_a^b$ of vectors are denoted by angle brackets 
$\psi^{<a>}\equiv h^a_b\psi^b$
and the projected symmetric trace-free (PSTF) of a tensor is given by
\begin{eqnarray}
\psi^{<ab>}&\equiv&\left[h^{(a}_c h^{b)}_d-\frac{1}{3}h^{ab}h_{cd}\right]\psi^{cd} \, .
\end{eqnarray}
The covariant time derivative and projected derivative are given by
\begin{eqnarray}
\dot \psi_{a..b}&\equiv&u^c\nabla_c\psi_{a...b}
\end{eqnarray}
and
\begin{eqnarray}
\tilde \nabla_c\psi_{a...b}&\equiv&h^f_c h^d_a...h^e_b\nabla_f\psi_{d...e}
\end{eqnarray}
respectively.
The covariant derivative of the 4-velocity, $u^a$, can be decomposed as
\begin{equation}
\nabla_a u_b=-u_a\dot u_b+\tilde\nabla_a u_b=-u_a\dot u_b+\frac{1}{3}\theta h_{ab}+\omega_{ab}+\sigma_{ab}
\end{equation}
where $\dot u_a\equiv u^b\nabla_b u_a$ is the acceleration, $\theta \equiv \tilde \nabla_a u^a$
the expansion, $\sigma_{ab}\equiv \tilde\nabla_{<a}u_{b>}$ the shear and 
$\omega_{ab}\equiv \tilde\nabla_{[a}u_{b]}$ the vorticity of $u^a$.

A formalism for a further split (1+2) with respect to a spatial vector $n^a$ (with $u^a n_a=0$) was
then developed in \cite{ClarksonBarrett, Clarkson}. Projections perpendicular to $n^a$ are made with
\begin{equation}
N_a^b=g_a^b+u_au^b-n_an^b \, .
\end{equation}
Projected vectors $v^{<a>}$ can be decomposed with respect to $n^a$ as
\begin{equation}
v^{<a>}=V n^a+V^a 
\end{equation}
with 
\begin{equation}
V\equiv n^a v_a  \quad
\hbox{and} 
\quad V^a\equiv N^{ab}v_b\equiv v^{\overline a} \, ,
\end{equation}
where a bar over an index denotes projection with $N^{ab}$.
Similarly PSTF tensors $\psi_{<ab>}$ can be decomposed as
\begin{eqnarray}\label{12decomp}
\psi_{<ab>}&=&\Psi \left(n_a n_ b- \frac{1}{2}N_{ab}\right)+2\Psi_{(a}n_{b)}+\Psi_{ab} \, ,
\end{eqnarray}
where
\begin{eqnarray}
\Psi &\equiv& n^a n^b \psi_{<ab>} \\
\Psi_a &\equiv& N_a^b n^c \psi_{<bc>} \\
\Psi^{ab}&\equiv&\left[N^{(a}_c N^{b)}_d-\frac{1}{2}N^{ab}N_{cd}\right]\psi^{<cd>}\equiv \Psi^{\{ab\}} \, .
\end{eqnarray}
Here $ \Psi^{\{ab\}}$ is symmetric and traceless.

Derivatives along and perpendicular to $n^a$ are given by
\begin{eqnarray}
\hat\psi_{a...b}&\equiv&n^c \tilde \nabla_c\psi_{a...b}=n^ch^f_c h^d_a...h^e_b\nabla_f\psi_{d...e}
\end{eqnarray}
and
\begin{eqnarray}
\delta_c\psi_{a...b}&\equiv&N_c^f N_a^d...N_b^e \tilde\nabla_f\psi_{d...e}
\end{eqnarray}
respectively.
Similarly to the derivative of the 4-velocity, the derivatives of $n_a$ can be decomposed as
\begin{equation}\label{eqspacen}
\tilde\nabla_a n_b =n_a a_b+\frac{1}{2}\phi N_{ab}+\xi\epsilon_{ab}+\zeta_{ab}
\end{equation}
and
\begin{equation}\label{eqdotn}
\dot n_a={\cal A}u_a+\alpha_a \, ,
\end{equation}
where
\begin{eqnarray}
\fl a_a\equiv \hat n_a , \quad \phi \equiv \delta_a n^a , \quad \xi\equiv\frac{1}{2}\epsilon^{ab}\delta_a n_b , \quad
\zeta_{ab}\equiv \delta_{\{a}n_{b\}} , \quad {\cal A}\equiv n^a\dot u_a , \quad \alpha_a\equiv \dot n_{\bar a}
\end{eqnarray}
and
\begin{equation}
\epsilon_{ab}\equiv\eta_{abc}n^c\equiv u^d\eta_{dabc}n^c .
\end{equation}
Here $\eta_{dabc}$ is the totally anti-symmetric 4-dimensional volume element with $\eta_{0123}=\sqrt{\vert \det g_{ab} \vert}$.
%%%%%%%%%%%%%%%%%%%%%%%%%%%%%%%%%%%%%%%%%%%%%%%%
\subsection{Fundamental equations in 1+3 split for the irrotational case}\label{section:13eq}
%%%%%%%%%%%%%%%%%%%%%%%%%%%%%%%%%%%%%%%%%%%%%%%%
The propagation and constraint equations are given in \cite{EllisvanElst}. We will here only
consider the case of a perfect fluid with vanishing vorticity. Imposing $\omega_{ab}=0$
introduces only one new constraint
\begin{equation}
\eta^{abc}\tilde\nabla_b\dot u_c=0 \, ,
\end{equation}
but with a barytropic equation of state, $p=p(\mu)$ where $p$ is pressure and $\mu$ energy density
in the rest frame of an observer, this equation is identically satisfied. From the Ricci identities one finds the 
following propagation equation for the expansion
\begin{equation}
\dot \theta-\tilde \nabla_a\dot u^a =-\frac{1}{3}\theta^2+\dot u_a \dot u^a-2\sigma^2-\frac{1}{2}(\mu+3p)+\Lambda \, ,
\end{equation}
where
\begin{equation}
\sigma^2\equiv\frac{1}{2}\sigma^{ab}\sigma_{ab} 
\end{equation}
and $\Lambda$ is the cosmological constant. 

The equation for the shear is 
\begin{equation}
\dot\sigma^{<ab>}-\tilde\nabla^{<a}\dot u^{b>}=-\frac{2}{3}\theta\sigma^{ab}+\dot u^{<a}\dot u^{b>}
-\sigma^{<a}\!_c\sigma^{b>c}-E^{ab} \, ,
\end{equation}
where $E_{ab}\equiv  C_{acbd}u^c u^d$ is the electric part of the Weyl tensor. 
\noindent
One also obtains the following constraints
\begin{equation}\label{divsigma}
\tilde\nabla_b\sigma^{ab}-\frac{2}{3}\tilde\nabla^a\theta=0 
\end{equation}
\begin{equation}
H^{ab}=({\rm{curl}}\, \sigma)^{ab}\equiv \eta^{cd<a}\tilde\nabla_c\sigma^{b>}\!_d \, ,
\end{equation}
where $H_{ab}\equiv \frac{1}{2}\eta_{ade}C^{de}\!\!_{bc}u^c$ is the magnetic part of the
Weyl tensor.

\noindent

From the twice contracted Bianchi identities one obtains
\begin{equation}
\dot\mu=-\theta(\mu+p)
\end{equation}

\begin{equation}
\tilde\nabla_a p+(\mu+p)\dot u_a=0
\end{equation}

\noindent
and the remaining Bianchi identities give the following propagation equations
\begin{equation}
\fl \dot E^{<ab>}=({\rm{curl}}\, H)^{ab}-\frac{1}{2}(\mu+p)\sigma^{ab}-\theta E^{ab}+
3\sigma^{<a}\!_c E^{b>c}+2\eta^{cd<a}\dot u_c H^{b>}\!_d
\end{equation}

\begin{equation}
\dot H^{<ab>}+({\rm{curl}}\, E)^{ab}=-\theta H^{ab}+3\sigma^{<a}\!_c H^{b>c}-
2\eta^{cd<a}\dot u_c E^{b>}\!_d
\end{equation}

\noindent
and constraints
\begin{equation}
\tilde\nabla_b E^{ab}-\frac{1}{3}\tilde\nabla^a\mu-\eta^{abc}\sigma_{bd}H^d\!_c=0
\end{equation}

\begin{equation}
\tilde\nabla_b H^{ab}+\eta^{abc}\sigma_{bd}E^d\!_c=0
\end{equation}
respectively.

By differentiating the constraints with respect to "time" and using the commutators between
the "time" and "spatial" derivatives it was shown in \cite{Maartens}
that the constraints are propagated for irrotational dust and
in \cite{vanElst} this result was extended to the barytropic case $p=p(\mu)$.
%%%%%%%%%%%%%%%%%%%%%%%%%%%%%%%%%%%%%%%%%%%%%%%%
\section{Kantowski-Sachs}\label{sec:KantowskiSachs}
%%%%%%%%%%%%%%%%%%%%%%%%%%%%%%%%%%%%%%%%%%%%%%%%
Kantowski-Sachs cosmologies, \cite{KantowskiSachs}, have a 4-dimensional isometry group acting multiply transitive
on 3-spaces with topology $R\times S_2$, i.e. they are locally rotationally symmetric (LRS). With zero vorticity the 
line-element can be written as
\begin{equation}
ds^2 = -dt^2 + a_1^2(t)dz^2+a_2^2(t)\left(d\vartheta^2+\sin^2\vartheta d\varphi^2\right)
\end{equation}
with the 4-velocity of matter given by $u=\frac{\partial}{\partial t}$ and the direction of anisotropy
by $n=\frac{1}{a_1}\frac{\partial}{\partial z}$. The expansion is given by
\begin{equation}\label{theta}
\theta = \frac{\dot a_1}{a_1}+2\frac{\dot a_2}{a_2}
\end{equation}
and in the tetrad  
\begin{eqnarray}\label{tetrad}
\omega^0=dt, \quad \omega^1=a_1dz, \quad
 \omega^2=a_2 d\vartheta, \quad \omega^3=a_2 \sin \vartheta d\varphi, 
\end{eqnarray}
the shear is given by
\begin{eqnarray}
\Sigma&\equiv& \sigma_{11}=-2\sigma_{22}=-2\sigma_{33} =\frac{2}{3}\left(\frac{\dot a_1}{a_1}
-\frac{\dot a_2}{a_2}\right)\, .
\end{eqnarray}

%%%%%%%%%%%%%%%%%%%%%%%%%%%%%%%%%%%%%%%%%%%%%%%%
\subsection{The evolution equations}
%%%%%%%%%%%%%%%%%%%%%%%%%%%%%%%%%%%%%%%%%%%%%%%%
Due to the LRS symmetry the shear and electric part of the Weyl tensor can be written as
\begin{equation}\label{sigma0}
\sigma_{ab}=\Sigma(n_a n_b-\frac{1}{2}N_{ab}) \, .
\end{equation}
and
\begin{equation}
E_{ab}={\cal E}(n_a n_b-\frac{1}{2}N_{ab})
\end{equation}
respectively in terms of the anisotropy vector $n^a$ and the projection operator $N_{ab}$. 
Given an equation of state and the cosmological constant, the Kantowski-Sachs models are completely
determined in terms of shear, $\Sigma$, expansion, $\theta$, and energy density, $\mu$.  
The electric part of the Weyl tensor is given algebraically as
\begin{eqnarray}
{\cal E}&=&- \frac{2}{3}\mu- \frac{2}{3}\Lambda-\Sigma^2+\frac{2}{9}\theta^2+\frac{1}{3}\Sigma\theta \, ,
\end{eqnarray}
whereas
\begin{equation}\label{gamKS2} 
H_{ab}=\dot u_a=a_a=\phi=\xi=\zeta_{ab}={\cal A}=\alpha_a=0 \, . 
\end{equation}
From the equations in section \ref{section:13eq} the following evolution equations are obtained
\begin{eqnarray}\label{eq01}
\dot \Sigma &=& \frac{2}{3}\mu+\frac{2}{3}\Lambda+\frac{1}{2}\Sigma^2-\Sigma\theta-\frac{2}{9}\theta^2 \\\label{eq02}
\dot \mu&=&-\theta (\mu + p) \\\label{eq03}
\dot \theta &=& -\frac{1}{3}\theta^2-\frac{1}{2}(\mu+3p-2\Lambda)-\frac{3}{2}\Sigma^2 \, .
\end{eqnarray}
Alternatively one of the equations can be replaced by
\begin{equation}\label{eq04}
\dot K=-(\frac{2}{3}\theta-\Sigma)K
\end{equation}
where $K$, given by
\begin{equation}\label{eqK}
K=\mu+\Lambda+\frac{3}{4}\Sigma^2- \frac{1}{3}\theta^2 > 0 \, ,
\end{equation}
is the curvature of the 2-spheres $S_2$.
%%%%%%%%%%%%%%%%%%%%%%%%%%%%%%%%%%%%%%%%%%%%%%%%
\subsection{Dynamical system analysis of background}\label{dynamical}
%%%%%%%%%%%%%%%%%%%%%%%%%%%%%%%%%%%%%%%%%%%%%%%%
Dynamical system analysis of the Kantowski-Sachs models with positive cosmological constant were done in, e.g. 
\cite{Weber, GoliathEllis}. We here follow the notation in \cite{GoliathEllis}.
The relations between their variables and ours are given by
\begin{eqnarray}\nonumber
D&=&\sqrt{\frac{1}{3}\mu+\frac{1}{3}\Lambda+\frac{1}{4}\Sigma^2}\, , \quad 
Q_0=\frac{\theta}{3D}\, , \\ 
Q_+&=&-\frac{\Sigma}{2D}\, ,
\quad \tilde\Omega_{\Lambda}=\frac{\Lambda}{3D^2}\, , \quad \Omega_D=\frac{\mu}{2D^2}\, .
\end{eqnarray}
The equilibrium points are given by
\vskip2mm
\noindent
$_{\pm}F$: Flat Friedmann: $\Lambda=\Sigma=0$, $\mu=\theta^2/3$, $Q_0=\pm 1$, saddle points

\noindent
$_+K_{\pm}$: Kasner: $\Lambda=\mu=0$, $\Sigma=\mp\frac{2}{3}\theta$, $Q_0=1$, $Q_+=\pm 1$, source points

\noindent
$_-K_{\pm}$: Kasner: $\Lambda=\mu=0$, $\Sigma=\pm\frac{2}{3}\theta$, $Q_0=-1$, $Q_+=\pm 1$, sink points

\noindent
$_{\pm}dS$: de Sitter: $\mu=\Sigma=0$, $\theta=\pm \sqrt{3\Lambda}$, sink/source points

\noindent
$_{\pm}X$: $\mu=0$, $\theta=\pm \sqrt{\Lambda}$, $\Sigma=\pm \frac{2}{3}\sqrt{\Lambda}$, saddle points.
\vskip2mm
\noindent
The points $_{\pm}X$ are exact vacuum solutions with metrics \cite{GoliathEllis}
\begin{equation}\label{Xmetrics}
ds^2=-dt^2+e^{\pm 2\sqrt{\Lambda}t}dz^2+\frac{1}{\Lambda}\left(d\vartheta^2+\sin^2\vartheta d\varphi^2\right) \, .
\end{equation}

%%%%%%%%%%%%%%%%%%%%%%%%%%%%%%%%%%%%%%%%%%%%%%%%
\subsection{Vacuum solutions}\label{sec:vacuum}
%%%%%%%%%%%%%%%%%%%%%%%%%%%%%%%%%%%%%%%%%%%%%%%%
In the vacuum case the equations can be integrated completely. If the curvature of $S_2$, $K$, is
a constant it follows that $\Sigma=\frac{2}{3}\theta$ (or $K=0$). The system (\ref{eq01})-(\ref{eq04}) 
then reduces to
$\dot\theta=\Lambda-\theta^2$
with solutions
\begin{equation}
\theta=\sqrt{\Lambda}\frac{Ce^{\sqrt{\Lambda}t}-e^{-\sqrt{\Lambda}t}}{Ce^{\sqrt{\Lambda}t}+e^{-\sqrt{\Lambda}t}} \, ,
\end{equation}
where $C$ is a constant of integration (and $\theta=\sqrt{\Lambda}$, corresponding to the critical point
$\!_+X$).
By redefining the time coordinate $t \rightarrow t+t_0$, these
solutions can be rewritten as one of the following two
\begin{equation}
\theta=\sqrt{\Lambda}\frac{\sinh(\sqrt{\Lambda}t)}{\cosh(\sqrt{\Lambda}t)}\, , \quad
\theta=\sqrt{\Lambda}\frac{\cosh(\sqrt{\Lambda}t)}{\sinh(\sqrt{\Lambda}t)} \, .
\end{equation}
($C=0$ gives the critical point $\!_-X$ with $\theta=-\sqrt{\Lambda}$.)
The corresponding line-element is given by
\begin{equation}
ds^2=-dt^2+f^2(t)dz^2+\frac{1}{\Lambda}\left(d\vartheta^2+\sin^2\vartheta d\varphi^2\right).
\end{equation}
where $f(t)$ for the first solution is given by
\begin{equation}\label{metricbounce}
f(t)=a_0\cosh(\sqrt{\Lambda}t)  \, .
\end{equation}
These spacetimes experience a bounce in the $z$-dirction at $t=0$ (and are non-expanding in the perpendicular
directions). They start at the critical point $\!_{-}X$ and end at $\!_{+}X$. For the other solution
one gets
\begin{equation}
f(t)=a_0\sinh(\sqrt{\Lambda}t)
\end{equation}
and hence these solutions are singular for $t=0$. They start at the critical point $\!_+K_-$ (Kasner) and ends
at $\!_+X$, or (for large negative $t$) start at $\!_-X$ and end at $\!_-K_+$.

If $K$ is not a constant it can be used as the independent coordinate. The system (\ref{eq01})-(\ref{eq04})
for $\mu=p=0$, together with the constraint (\ref{eqK}),
then reduces to (see \cite{MarklundBradley} for details)
\begin{eqnarray}
\fl \theta=\pm \frac{2\Lambda-4K+6MK^{3/2}}{2\sqrt{2MK^{3/2}-K+\Lambda/3}}\, ,\quad  
\Sigma=\pm \frac{2K-6MK^{3/2}}{3\sqrt{2MK^{3/2}-K+\Lambda/3}} \, ,
\end{eqnarray}
where $M$ is a constant of integration.
Changing independent coordinate through $K=1/T^2$ the corresponding line-element becomes
\begin{eqnarray}
\fl ds^2=-\frac{dT^2}{\left(\frac{2M}{T}-1+\frac{\Lambda}{3}T^2\right)}+\left(\frac{2M}{T}-
1+\frac{\Lambda}{3}T^2\right)dz^2+ 
T^2\left(d\vartheta^2+\sin^2\vartheta d\varphi^2\right) \, .
\end{eqnarray}
If $A\equiv 2M/T-1+\Lambda T^2/3>0$ this is the space-homogeneous region of the Schwarzschild-de Sitter
metric \cite{StuchlikHledik}, and $M=0$ gives de Sitter space. With $\Lambda M^2/3>1/27$ the requirement
$A>0$ is satisfied for all positive $T$ if $M>0$ and
the metric starts at the critical point $\!_+K_+$ ($T=0$) and ends at $\!_+ dS$ (de Sitter)
 .  $A>0$ also in the region $(-\infty,T_0)$, where $T_0<0$
is the only real solution of $A=0$. These spacetimes start at $\!_-dS$ and ends at $\!_-K_+$.
If $M<0$ the situation is reversed so that for $T<0$ there is a class of solutions starting at $\!_-dS$ for
large negative $T$ and ending at $\!_-K_-$ for $T=0$, and another class starting at $\!_+K_-$
for a positive $T_0$ and ending at $\!_+dS$ for large positive $T$. 

For $0<\Lambda M^2/3<1/27$ and $M>0$ there is one class of solutions starting at $\!_-dS$ at negative
infinity and ending at $\!_-K_+$ for a negative $T_0$, another class starting at $\!_+K_+$ for $T=0$
and ending at $\!_-K_+$ for a positive $T_1$ and a third class starting at $\!_+K_-$ for a positive
$T_2$ and ending at $\!_+dS$ for large positive $T$. Similarly, for negative $M$, there is one class
going from $\!_-dS$ to $\!_-K_+$, a second from $\!_+K_-$ to $\!_-K_-$ and a third from $\!_+K_-$ to $\!_+dS$.
%%%%%%%%%%%%%%%%%%%%%%%%%%%%%%%%%%%%%%%%%%%%%%%%
\subsection{Solutions with matter}
%%%%%%%%%%%%%%%%%%%%%%%%%%%%%%%%%%%%%%%%%%%%%%%%
The general dust solutions in terms of elliptic functions were found in \cite{Lorenz1,Lorenz2}. A few
of them can be given in terms of elementary functions, see e.g. \cite{Lorenz2,GronEriksen}. For certain
choices of the parameters in \cite{Lorenz2} physically reasonably solutions starting from a pancake like 
singularity can be obtained. These solutions end in expanding de Sitter.
The solution \cite{GronEriksen} starts in $_+F$ and ends in $_+X$.

We here consider approximate solutions with 
$\mu\sim p \ll \Lambda$. To first order the dependent quantities are written as $\theta=\theta_0+\theta_1$,
$\Sigma=\Sigma_0+\Sigma_1$ and $\mu=\mu_1$. Assuming a linear equation of state $p=(\gamma-1)\mu$ 
and $\Sigma_0=\frac{2}{3}\theta_0$ for the background vacuum metric, the following first order system
\begin{eqnarray}\label{approx1}
\dot \mu_1&=&-\theta_0\gamma\mu_1\\
\dot \Sigma_1-\frac{2}{3}\dot \theta_1&=&\gamma\mu_1+\theta_0\left(\Sigma_1-\frac{2}{3}\theta_1\right)\\\label{approx3}
\dot\Sigma_1+\frac{5}{6}\dot\theta_1&=&\frac{1}{4}(6-5\gamma)\mu_1
-2\theta_0\left(\Sigma_1+\frac{5}{6}\theta_1\right)
\end{eqnarray}
is obtained. In appendix \ref{appA} solutions around the bounce metric (\ref{metricbounce}) and the 
critical points $\!_{\pm}X$, (\ref{Xmetrics}), are given.
Since for all of these solutions some terms grow unbounded, they are only valid for limited
time intervals. Nevertheless they can be used to check our numerical codes for shorter
time intervals in the case with densities that are initially low, and also for finding suitable starting conditions.

%%%%%%%%%%%%%%%%%%%%%%%%%%%%%%%%%%%%%%%%%%%%%%%%
\subsection{Bouncing solutions}\label{sectionbounce}
%%%%%%%%%%%%%%%%%%%%%%%%%%%%%%%%%%%%%%%%%%%%%%%%
Bouncing FLRW universes containing dust and radiation with cosmological constant are
excluded by observations \cite{Ehlers}, see also \ref{appE}. This relies on a very simple
argument. If quasars are observed at redshift $z>4$, the dust component $%
\Omega _{m}$ cannot overcome the value $0.02$, otherwise the bounce would
occur at a smaller redshift than $4$. We now consider whether a similar argument 
also holds for bouncing Kantowski-Sachs universes with
cosmological constant. The bounce occurs either

\noindent
$i)$ in the
direction of anisotropy at $a_{1\ast }$ for which $\dot{a}_{1\ast }=0$ and $%
\ddot{a}_{1\ast }>0$ 

\noindent
or 

\noindent
$ii)$ in the perpendicular direction at $a_{2\star }
$ for which $\dot{a}_{2\star }=0$ and $\ddot{a}_{2\star }>0$.

By defining the average scale factor $a$ through
\begin{equation}\label{average}
\theta=3\frac{\dot a}{a} \, .
\end{equation}
and using (\ref{theta}) one obtains $a^3=a_1a^2$.
If we assume that the universe is filled with dust ($\gamma=1$) and radiation ($\gamma=4/3$) Eq. (\ref{eq02}) is satisfied by the following 
density 
\begin{equation}
\mu=\mu_{m}+\mu_{r}=\mu_{m0}\left(\frac{a_0}{a}\right)^3+
\mu_{r0}\left(\frac{a_0}{a}\right)^4 \, ,
\end{equation}
where the subscript 0 denotes present values.
We then introduce the usual dimensionless parameters
\begin{equation}\label{Omegas}
\Omega _{m} =3\mu _{m0}/\theta _{0}^{2}\ , \quad
\Omega _{r} =3\mu _{r0}/\theta _{0}^{2}\     \quad \hbox{and} \quad
\Omega _{\Lambda } =3\Lambda /\theta _{0}^{2} \, .
\end{equation}

\noindent
Case $i)$: Since $\theta /3+\Sigma =\dot{a}_{1}/a_{1}$, we have $0<%
\dot{\theta}_{\ast }/3+\dot{\Sigma}_{\ast }$ and by using Eqs. (\ref{eq01}), (\ref{eq03})
and the notations (\ref{Omegas}) this leads to%
\begin{equation}
0<\frac{3}{\theta_0^2}\frac{\ddot a_{1*}}{a_{1*}}=\frac{\Omega _{m}}{2}\left( \frac{a_{0}}{a_\ast}\right) ^{3}+
\frac{\Omega _{r}}{3}\left( \frac{a_{0}}{a_\ast}\right) ^{4}+
\Omega _{\Lambda
}\ .
\end{equation}%
Eq. (\ref{eqK}) gives
\begin{equation}
0<\frac{3}{\theta_0^2a_{2*}^2}+\frac{3}{\theta_0^2}\frac{\dot a_{2*}^2}{a_{2*}^2}=\Omega_m\left(\frac{a_0}{a_*}\right)^3+
\Omega_{\Lambda}
\end{equation}
and another
combination of Eqs. (\ref{eq01}) and (\ref{eq03}), $2\dot{\theta}_{\ast }/3-\dot{%
\Sigma}_{\ast }$, gives
\begin{equation}
-\frac{6}{\theta_0^2}\frac{\ddot a_{2*}}{a_{2*}}=\Omega_m\left(\frac{a_0}{a_*}\right)^3
+\frac{4}{3}\Omega_r\left(\frac{a_0}{a_*}\right)^4 \, .
\end{equation}
The inequalities are always satisfied with a positive cosmological constant and the equations also give
the values of $\dot{a}_{2\ast }$, $\ddot{a}_{2\ast }$ and $\ddot{a}_{1\ast }$ at the
bounce. Therefore, we have no other constraints unless we can find some further integrals to the
system (\ref{eq01})-(\ref{eq03}).

We can also consider the equations (\ref{eq01}), (\ref{eq03}) and (\ref{eqK}) at the present time $t_0$.
Neglecting radiation one obtains
\begin{eqnarray}\label{OLambda}
-\frac{1}{3}q_{10}&=&\frac{1}{2}\Omega_m+\Omega_{\Lambda}-
\frac{2H_{10}H_{20}}{3H_0^2}%\frac{\dot a_{10}}{a_{10}}\frac{\dot a_{20}}{a_{20}}
\\
-\frac{2}{3}q_{20}&=&-\Omega_m
+\frac{2H_{10}H_{20}}{3H_0^2}%\frac{\dot a_{10}}{a_{10}}\frac{\dot a_{20}}{a_{20}}
\\\label{2curv}
\frac{1}{3H_0^2a_{20}^2}&=&\Omega_m+\Omega_{\Lambda}-1+\frac{\left(\Delta H_0\right)^2}{9H_0^2} \, ,
\end{eqnarray}
where  $H_0= \dot a_0/a_0$ is the present value of the Hubble constant,  $H_{10}= \dot a_{10}/a_{10}$,
 $H_{20}= \dot a_{20}/a_{20}$, $\Delta H_0=H_{10}-H_{20}$, $q_{10}=-\ddot a_{10}/(a_{10}H_0^2)$
and  $q_{20}=-\ddot a_{20}/(a_{20}H_0^2)$.
How the quantities $H_{10}$,  $H_{20}$,  $q_{10}$ and  $q_{20}$ relate to observations 
can be seen from the general expression for redshift (when both emitter ($E$) and receiver ($R$) have fixed
spatial coordinates)
\begin{equation}
1+z=\frac{\nu _{E}}{\nu _{R}}=\frac{p_{t}\left( E\right) }{p_{t}\left(
R\right) }\sqrt{\frac{g_{tt}\left( R\right) }{g_{tt}\left( E\right) }}\ 
\end{equation}
 (see for example \cite{Hobson}).
Here $p_{a}=g_{ab}dx^{b}/d\lambda $ (with affine parameter $\lambda $)
denotes the covariant components of photon 4-momentum. By integrating the geodesic equations for the Kantowski-Sachs
metric the following
redshift formula is obtained for photons moving in the $\theta=\pi/2$ plane
\begin{equation}
1+z=\frac{a_{0}}{a}\left( \frac{\alpha _{0}}{\alpha }\right) ^{2/3}\left[ 
\frac{p_{z}^{2}+p_{\varphi }^{2} \alpha ^{2}}{%
p_{z}^{2}+p_{\varphi }^{2} \alpha _{0}^{2}}%
\right]^{1/2} \ ,  \label{redshift}
\end{equation}
where $\alpha =a_{1}/a_{2}$  and $p_z$ and $p_\varphi$ are integration constants.  
For example, photons moving in the $z$-direction have a redshift $1+z=a_{10}/a_1$ whereas those moving
along the 2-spheres have redshifts $1+z=a_{20}/a_2$. 

In the isotropic limit $\Delta H_0 \rightarrow 0$, 
$\Delta q_0=q_{10}-q_{20} \rightarrow 0$ 
and $1/a_{20}^2 \rightarrow 0$ the equations (\ref{OLambda})-(\ref{2curv}) agree with those of the flat FLRW model with
cosmological constant 
($q_0\equiv \ddot a_0 a_0/\dot a_0^2=\Omega_m/2-\Omega_{\Lambda}$, $1=\Omega_m+\Omega_{\Lambda}$).
Since there are indications of anistropies, but with large uncertainties,  both in the Hubble and deceleration parameters,
see e.g. \cite{McClureDyer} and \cite{CaiTuo}, it would be of interest to get better estimates
of these parameters, even if present values do not seem to support Kantowski-Sachs models.

\noindent
Case $ii)$: From $K=1/a_2^2$ we find $\ddot{K}=2\left( -\ddot{a}_{2}/a_{2}+3\dot{a}%
_{2}^{2}/a_{2}^{2}\right) K$ and therefore $\ddot{K}_{\star }<0$. Then, taking the derivative of 
Eq. (\ref{eq04}) to get another expression for $\ddot K$ and using Eqs. (\ref{eq01}) and (\ref{eq03}) together with $\theta
_{\star }=3\Sigma _{\star }/2$, we find  
\begin{equation}
\Omega _{m}<0\ .
\end{equation}%
Hence a bounce in the scale factor $a_{2}$ is not possible, whereas a bounce in the
direction of anistropy cannot be excluded simply from this type of argument. 

If we instead look at a bounce in the average scale factor at $a=a_\#$, so that 
$\theta_\#=3\dot a_\#/a_\#=0$ and $\ddot a_\# > 0$, the following inequality
\begin{equation}
\frac{2}{3H_0^2}\frac{1}{a_{2\#}^2}+\Omega _{\Lambda }
<\frac{3}{2}\Omega _{m}\left( \frac{a_{0}}{a_\#}\right) ^{3}\ 
\end{equation}
can be derived, but no upper bound for the redshift
at the bounce can be obtained from this either.

%%%%%%%%%%%%%%%%%%%%%%%%%%%%%%%%%%%%%%%%%%%%%%%%
\section{Perturbations on Kantowski-Sachs}\label{sec:pert}
 %%%%%%%%%%%%%%%%%%%%%%%%%%%%%%%%%%%%%%%%%%%%%%%%
In this section we calculate the equations governing the growth of density perturbations on a Kantowski-Sachs background 
to first order.  The inhomogeneities will be described by quantities that are zero on the background,
and hence are gauge invariant \cite{StewartWalker}.
The primary variable is the density gradient 
\begin{eqnarray}
{\cal D}_a &\equiv& \frac{a\tilde \nabla_a \mu}{\mu} \, .
\end{eqnarray}
Here $a$ is the average scale factor, defined in (\ref{average}). 
%\begin{equation}
%\theta=3\frac{\dot a}{a} \, .
%\end{equation}
The density fluctuations $\frac{\delta \mu}{\mu}$ on a length scale $l$
are related to the quantity ${\cal D}_a$ through $\frac{\delta \mu}{\mu}\sim
({\cal D}_a{\cal D}^a)^{1/2} l/a=({\cal D}_a{\cal D}^a)^{1/2} l_0$, where
$l_0=l/a$ is the comoving dimensionless length scale. However, note that the
quantity 
\begin{equation}
\delta(x)\equiv \frac{\mu(x)-\bar\mu(x)}{\mu(x)}
\end{equation}
depends on the identification between the fictitious background with density $\bar\mu(x)$
and the real universe with density $\mu(x)$ and can be given given any value by changing
the identification \cite{EllisvanElst}, whereas ${\cal D}_a$ is gauge invariant.

To close the system three auxiliary quantities, that we choose as 
\begin{eqnarray}\label{defZ}
{\cal Z}_a \equiv a\tilde \nabla_a\theta \, ,\quad
{\cal T}_a \equiv a\tilde \nabla_a\sigma^2 \quad \hbox{and}
\quad {\cal S}_a \equiv a\tilde \nabla_a(\sigma^{ab}S_{ab}) \, 
\end{eqnarray}
will be needed. 
Here the traceless part of the 3-Ricci tensor is given by
\begin{eqnarray}\nonumber
S_{ab}&\equiv& \!^3R_{ab}-\frac{1}{3}\,\, \!^3R h_{ab}=-\dot\sigma_{<ab>}-\theta\sigma_{ab}
+\tilde\nabla_{<a}\dot u_{b>}+\dot u_{<a}\dot u_{b>}\\
&=&E_{ab}+\sigma_{<a}\!^c\sigma_{b>c}-\frac{1}{3}\theta\sigma_{ab} \, .
\end{eqnarray}

In accordance with (\ref{12decomp}) $\sigma_{ab}$ and $S_{ab}$ can be decomposed as 
\begin{equation}
\sigma_{ab}=\Sigma(n_a n_b-\frac{1}{2}N_{ab})+2\Sigma_{(a}n_{b)}+\Sigma_{ab}
\end{equation}
and 
\begin{equation}\label{Stilde}
S_{ab}=\tilde S(n_a n_b-\frac{1}{2}N_{ab})+2\tilde S_{(a}n_{b)}+\tilde S_{ab} \,
\end{equation}
respectively.
Note that $S_{ab}$ to zeroth order is given in terms of the other quantities, but to first order
is an independent quantity.
%%%%%%%%%%%%%%%%%%%%%%%%%%%%%%%%%%%%%%%%%%%%%%%%
\subsection{The first order equations}
%%%%%%%%%%%%%%%%%%%%%%%%%%%%%%%%%%%%%%%%%%%%%%%%
The propagation equations for the gradients are obtained by taking the gradients $\tilde\nabla_a$
of the propagation equations in section \ref{section:13eq} and then using the commutator 
between
"time" and "spatial" derivatives acting on a scalar, that to first order reduces to \cite{EllisvanElst}
\begin{equation}
\tilde\nabla_a(\dot f)-(\tilde\nabla_a f)^.=-\dot u_a \dot f +\frac{1}{3}\theta\tilde \nabla_a f 
+ \sigma_a\,^c \tilde \nabla_c f \, .
\end{equation}
In \ref{appB} some details of the calculations are given. Finally the following
first order system
\begin{eqnarray}\label{eqD}
\fl \dot {\cal D}_a =\frac{\theta p}{\mu}{\cal D}_a-\frac{3}{2}\Sigma n_a n^c {\cal D}_c+
\frac{1}{2}\Sigma{\cal D}_a-{\cal Z}_a(1+\frac{p}{\mu}) \\ \nonumber
\fl \dot {\cal Z}_a =-\frac{2}{3}\theta {\cal Z}_a-\frac{1}{2}\mu{\cal D}_a-\frac{3}{2}\Sigma n_a n^c {\cal Z}_c+
\frac{1}{2}\Sigma {\cal Z}_a-2{\cal T}_a+ 
\frac{3}{2}\frac{\mu p'}{\mu+p}\left(\tilde S+\frac{3}{2}\Sigma^2\right){\cal D}_a-\\\label{eqZ}
\frac{\mu p'}{\mu+p}\tilde\nabla_a\tilde\nabla^b{\cal D}_b \\ \nonumber
\fl \dot {\cal T}_a = -2\theta{\cal T}_a-\frac{3}{2}\Sigma n_a n^c{\cal T}_c+\frac{1}{2}\Sigma{\cal T}_a-
{\cal S}_a-\frac{3}{2}\Sigma^2{\cal Z}_a+ 
\frac{3}{2}\Sigma\left(\tilde S + \theta\Sigma\right)
\frac{\mu p'}{\mu + p}{\cal D}_a- \\\label{eqT}
\fl \frac{\mu p'}{\mu + p}\Sigma(\frac{3}{2}n_c n^b \tilde\nabla_a \tilde\nabla^c{\cal D}_b-
\frac{1}{2}\tilde\nabla_a\tilde\nabla^b{\cal D}_b) \\\nonumber
\fl \dot {\cal S}_a=\left(\Sigma^2+2\frac{\tilde S^2}{\Sigma^2}\right){\cal T}_a+\left(\frac{5}{2}\Sigma-
\frac{5}{3}\theta-
2\frac{\tilde S}{\Sigma}\right){\cal S}_a- 
\frac{3}{2}\Sigma n_a n^c {\cal S}_c-
 \Sigma\left(\frac{5}{2}\tilde S +\frac{2}{3}\Sigma\theta\right){\cal Z}_a+\\\nonumber
\fl \frac{p'\mu}{\mu+p}\tilde S\left(\frac{5}{2}\theta\Sigma+\frac{3}{2}{\tilde S}-\frac{3}{2}\Sigma^2
\right){\cal D}_a+\mu\Sigma^2 {\cal D}_a+
\frac{p'\mu}{\mu+p}\left[\frac{1}{2}\left({\tilde S}-
\frac{1}{3}\theta\Sigma+2\Sigma^2\right)\tilde\nabla_a\tilde\nabla^b{\cal D}_b-\right. \\\label{eqS}
\fl \left. \frac{3}{2}\left(\tilde S-\frac{1}{3}\theta\Sigma+\Sigma^2\right)n^b n^c\tilde\nabla_a\tilde\nabla_b{\cal D}_c
\right]+
\frac{3}{2}\Sigma n^b n^c\tilde\nabla_a\tilde\nabla_b{\cal Z}_c-
\frac{1}{2}\Sigma\tilde\nabla_a\tilde\nabla^b{\cal Z}_b-\tilde\nabla_a\tilde\nabla^b{\cal T}_b \, ,
\end{eqnarray}
where
\begin{equation}
\tilde S=-\frac{2}{3}\mu-\frac{2}{3}\Lambda-\frac{1}{2}\Sigma^2+\frac{2}{9}\theta^2=-\frac{2}{3}K<0
\end{equation}
to zeroth order and $p'\equiv dp/d\mu$, is obtained. As seen two apparently singular terms,  $2\tilde S^2/\Sigma^2 {\cal T}_a$ and
$-2\tilde S/\Sigma {\cal S}_a$, appear in (\ref{eqS}). Near points
where $\Sigma=0$ it is hence suitable to remove these terms by changing the dependent variable ${\cal T}_a$
to $\tilde {\cal T}_a$ through
\begin{equation}
{\cal T}_a=\Sigma^2 \tilde {\cal T}_a+\frac{\Sigma}{\tilde S}{\cal S}_a \, .
\end{equation}
Instead factors $1/\tilde S$ and $1/\tilde S^2$ will be introduced, but since
$\tilde S=-2K/3<0$ these will be well behaved. 
%%%%%%%%%%%%%%%%%%%%%%%%%%%%%%%%%%%%%%%%%%%%%%%%
\subsection{The projected equations}
%%%%%%%%%%%%%%%%%%%%%%%%%%%%%%%%%%%%%%%%%%%%%%%%
The equations (\ref{eqD})-(\ref{eqS}) can be decomposed into two sets by projecting with $n_a$ and $N_{ab}$
respectively.
For Kantowski-Sachs, for which (\ref{gamKS2}) holds, it follows from equations (\ref{eqdotn}) and (\ref{eqspacen}) 
for the
derivatives $\dot n_a$ and $\tilde\nabla_a n_b$ of $n_a$ that
\begin{equation}
\dot n_a =\tilde\nabla_a n_b=0 \, .
\end{equation}
Since $\dot u^a=0$ and $h_{ab}u^a=0$ it also follows that the derivatives of $N_{ab}$, $\dot N_{ab}$
and $\tilde\nabla_c N_{ab}$, become zero:
\begin{displaymath}
\dot N_{ab}=u^c \nabla_c(g_{ab}+u_a u_b+n_a n_b)=\dot u_a u_b+\dot u_b u_a=0
\end{displaymath}
\begin{displaymath}
\tilde\nabla_c N_{ab}=\tilde \nabla_c(u_a u_b)=h_c\!^fh_a\!^dh_b\!^e(u_d\nabla_fu_e+u_e\nabla_fu_d)=0 .
\end{displaymath}
Hence, since we only need $n_a$ and $N_{ab}$ to zeroth order, we can just let $n^a$ and $N_{ab}$
"pass through" the derivatives when projecting equations (\ref{eqD})-(\ref{eqS}).

With the definitions 
\begin{equation}
{\cal D}\equiv{\cal D}_an^a \, , \; {\cal Z}\equiv{\cal Z}_an^a \, , \; {\cal T}\equiv{\cal T}_an^a \, , \; 
{\cal S}\equiv{\cal S}_an^a 
\end{equation}
the equations projected along $n^a$ become:
\begin{eqnarray}\label{ds}
\fl \dot {\cal D} &=&\left(\frac{\theta p}{\mu}-\Sigma\right) {\cal D}-(1+\frac{p}{\mu}){\cal Z} \\ \label{zs} 
\fl \dot {\cal Z} &=&-\frac{1}{2}\mu{\cal D}-\left(\frac{2}{3}\theta+\Sigma \right) {\cal Z}-2{\cal T}+
\frac{3}{2}\frac{\mu p'}{\mu+p}\left(\tilde S+\frac{3}{2}\Sigma^2\right){\cal D}
-\frac{\mu p'}{\mu+p}n^a\tilde\nabla_a\tilde\nabla^b{\cal D}_b \\ \nonumber
\fl \dot {\cal T} &=& -\left(2\theta+\Sigma \right){\cal T}-\frac{3}{2}\Sigma^2{\cal Z}+ 
\frac{3}{2}\frac{\mu p'}{\mu + p}\Sigma\left(\tilde S + \theta\Sigma\right){\cal D}-  
{\cal S}-\frac{\mu p'}{\mu + p}\Sigma(\frac{3}{2}  n^a\tilde\nabla_an^b \tilde\nabla_b{\cal D}-\\\label{ts}
\fl && \frac{1}{2}n^a\tilde\nabla_a\tilde\nabla^b{\cal D}_b) \\\nonumber
\fl \dot {\cal S}&=&\left(\Sigma^2+2\frac{\tilde S^2}{\Sigma^2}\right){\cal T}+\left(\Sigma-\frac{5}{3}\theta-
2\frac{\tilde S}{\Sigma}\right){\cal S}-
\Sigma\left(\frac{5}{2}\tilde S +
\frac{2}{3}\Sigma\theta\right){\cal Z}+
\frac{p'\mu}{\mu+p}\tilde S\left(\frac{5}{2}\theta\Sigma+\frac{3}{2}{\tilde S}-\frac{3}{2}\Sigma^2
\right){\cal D}+\\\nonumber
\fl &&\mu\Sigma^2 {\cal D}+ \frac{p'\mu}{\mu+p}\left[\frac{1}{2}\left({\tilde S}-
\frac{1}{3}\theta\Sigma+2\Sigma^2\right)n^a\tilde\nabla_a\tilde\nabla^b{\cal D}_b - \right. 
\left.
\frac{3}{2}\left(\tilde S-\frac{1}{3}\theta\Sigma+\Sigma^2\right)n^a \tilde\nabla_an^b\tilde\nabla_b{\cal D}\right]-
\\ \label{ss}
\fl && n^a\tilde\nabla_a\tilde\nabla^b{\cal T}_b 
+\frac{3}{2}\Sigma n^a  \tilde\nabla_an^b\tilde\nabla_b{\cal Z}-
\frac{1}{2}\Sigma n^a\tilde\nabla_a\tilde\nabla^b{\cal Z}_b .
\end{eqnarray}
The terms $\tilde\nabla^b {\cal D}_b$ etc. can be decomposed as
\begin{equation}
\tilde\nabla^b{\cal D}_b=n^b\tilde\nabla_b{\cal D}+\delta^b {\cal D}_{\bar b}
\end{equation}
to first order.

The orthogonal equations, obtained by projecting with $N_{ab}$, can be found in \ref{appC}. 

%%%%%%%%%%%%%%%%%%%%%%%%%%%%%%%%%%%%%%%%%%%%%%%%
\subsection{Scalar equations}
%%%%%%%%%%%%%%%%%%%%%%%%%%%%%%%%%%%%%%%%%%%%%%%%
To treat the spatial derivatives appearing in the equations we will do a harmonic decomposition. For this purpose 
it is suitable to get the spatial derivatives in the form of two Laplace-like operators, 
\begin{equation}
\delta^2 \equiv \delta_a \delta^a \quad \hbox{and} \quad \hat \Delta \equiv n^a \tilde \nabla_a n^b \tilde\nabla_b 
\end{equation}
acting on our variables. To obtain this we define new variables
\begin{equation}\label{newvar}
\hat {\cal D} \equiv n^a\tilde\nabla_a{\cal D} \quad \hbox{and} \quad \slacal{D}\equiv \delta^a{\cal D}_{\bar a}
\end{equation}
and similarly for the other variables.
We then act on the system  (\ref{ds})-(\ref{ss})   with the operator $n^a\tilde\nabla_a$ and use the commutation relation 
\begin{equation}
\hat {\dot\Psi } -\dot {\hat \Psi }=\left(\frac{1}{3}\theta+\Sigma\right) \hat\Psi \, ,
\end{equation}
\cite{Clarkson}, that holds to first order. 
To remove the singular terms $2\tilde S^2/\Sigma^2 {\hatcal T}$ and $-2\tilde S/\Sigma {\hatcal S}$
we now also make the aforementioned change of the dependent variables ${\hatcal T}$ and $\slacal T$
to $\tilde {\hatcal T}$ and $\tilde {\slacal T}$ through
\begin{equation}
{\hatcal T}=\Sigma^2 {\tilde {\hatcal T}}+\frac{\Sigma}{\tilde S}{\hatcal S} 
\quad \hbox{and} 
\quad
{\slacal T}=\Sigma^2 {\tilde {\slacal T}}+\frac{\Sigma}{\tilde S}{\slacal S} 
\end{equation}
respectively.
The system for the hat variables then becomes
\begin{eqnarray}\label{eqhat1}
\fl \dot {\hat{\cal D} } &=& \bras{ \theta \bra{\frac{ p}{\mu} - \frac{1}{3}} - 2\Sigma } \hat{\cal D} - 
\bra{1+\frac{p}{\mu}} \hat{\cal Z} \\ 
\fl \dot {\hat{\cal Z} } &=&- \bra{\theta + 2 \Sigma  } \hat{\cal Z}  +
\bras{ -\frac{1}{2}\mu + \frac{3}{2}\frac{\mu p'}{\mu + p}\bra{\tilde S + \frac{3}{2}\Sigma^2} } \hat{\cal D} 
 -2\frac{\Sigma}{\tilde S}\hatcal S- 2 \Sigma^2{\tilde{\hatcal T}}
 - \frac{\mu p'}{\mu + p}\hat \Delta \bras{  \hat{\cal D} +  \slacal{D}} \\\nonumber
\fl \dot {\tilde{\hatcal T}}  &=& -\bra{\frac{1}{3}\theta + 2 \Sigma+\frac{\Sigma^3}{\tilde S}}{\tilde{ \hatcal T}}- 
\left(\frac{\Sigma^2}{\tilde S^2}+\frac{1}{\tilde S}\right)\hatcal S - 
\left[\frac{\Sigma \mu}{\tilde S}+\frac{\mu p'}{\mu + p} \left(\theta-\frac{3}{2}\Sigma\right)\right]\hatcal D 
+\left(1+ \frac{2}{3}\frac{ \Sigma\theta}{\tilde S}\right) \hatcal Z + \\ \nonumber
\fl  && \frac{\mu p'}{\mu + p}\frac{1}{\tilde S} 
\bras{\left(\frac{1}{2}\Sigma-\frac{1}{3}\theta\right)\hat \Delta \hatcal D -\left(\Sigma- \frac{1}{6}\theta\right) 
\hat \Delta \bra{\slacal{D}} }- 
\frac{1}{\tilde S}\hat\Delta(\hatcal Z-\frac{1}{2}\slacal Z)
+\frac{\Sigma}{\tilde S}\hat\Delta({\tilde{\hatcal T}}+{\tilde{\slacal T}})+\\
\fl && \frac{1}{\tilde S^2}\hat \Delta
(\hatcal S+ \slacal S)
\\ \nonumber
\fl \dot {\hatcal S} &=& \bras{ \mu \Sigma^2 +
\frac{\mu p'}{\mu + p} \tilde S \bra{ \frac{5}{2}\theta \Sigma + \frac{3}{2} \tilde S - \frac{3}{2} \Sigma^2 } } 
\hatcal D -
\bra{\frac{2}{3} \theta \Sigma + \frac{5}{2} \tilde S } \Sigma \hatcal Z   
  + \bra{\Sigma^4 + 2 \tilde S^2}{ \tilde{\hatcal T}} + 
\\ \nonumber
\fl && \bra{\frac{\Sigma^3}{\tilde S}-2\theta  } \hatcal S
+ \Sigma \hat \Delta \left(\hatcal Z - 
\frac{1}{2} \slacal{Z} \right) 
 -\Sigma^2 \hat \Delta \left( {\tilde{\hatcal T}}+\tilde{\slacal{T}}\right)	 
+	\frac{\mu p'}{\mu + p}\left[\bra{-\tilde S + \frac{1}{3} \theta \Sigma - \frac{1}{2} \Sigma^2} \hat \Delta 
\hatcal D
\right. \\ 
\fl && \left. 
+\frac{1}{2}\bra{\tilde S - \frac{1}{3}\theta \Sigma + 2 \Sigma^2  }\hat \Delta \bra{\slacal{D}}\right]  
-\frac{\Sigma}{\tilde S}\hat\Delta(\hatcal S+\slacal S) .
\label{eqhat4}
\end{eqnarray}
By taking the 2-divergence of the system (\ref{dv})-(\ref{sv}) and using the the commutation relation
\begin{equation}
\delta^a \dot \Psi_{\bar a}- \left(\delta^a \Psi_{\bar a}\right)^.=\left(\frac{1}{3}\theta-
\frac{1}{2}\Sigma\right)\delta^a \Psi_{\bar a} \, ,
\end{equation}
\cite{Clarkson},
(where $\delta^a \Psi_{\bar a}=\sla \Psi $ according to the above definition)
a similar system, that can be found in \ref{appC}, is obtained for the slashed variables. 
As we will see in the next section, the hat and slash variables are closely related.

Since the scale factors $a_1(t)$ and $a_2(t)$ appear in the spatial derivatives, the time
dependence of the variables $\hatcal D$ and $\slacal D$ will go as ${\cal D}/a_1$ and
${\cal D}_{\bar a}/a_2$ respectively. This is most easily seen by calculating $\hatcal D$
and $\slacal D$ in the tetrad (\ref{tetrad}), giving 
\begin{equation}
{\hatcal D}=\frac{1}{a_1}\frac{\partial {\cal D}}{\partial z} \quad \hbox{and} \quad
{\slacal D}=\frac{1}{a_2}\left(\frac{\partial D_{\bar 2}}{\partial \vartheta}+\frac{1}{\sin\vartheta}
\frac{\partial D_{\bar 3}}{\partial \varphi}\right) .
\end{equation}
Hence the time dependences of the variables 
\begin{equation}\label{DpDp}
{\cal D}_{\parallel} = a_1 \hatcal D \quad \hbox{and} \quad {\cal D}_{\perp} = a_2 \slacal D
\end{equation}
give a better representation 
of the development of
the relative density contrast $\frac{\delta \mu}{\mu}$.

%%%%%%%%%%%%%%%%%%%%%%%%%%%%%%%%%%%%%%%%%%%%%%%%
\subsection{Harmonic decomposition}
%%%%%%%%%%%%%%%%%%%%%%%%%%%%%%%%%%%%%%%%%%%%%%%%
We will use a harmonic decomposition
\begin{equation}\label{decomp}
\Psi=\sum\limits_{k_{\parallel}, k_{\perp}}\Psi_{k_{\parallel} k_{\perp}}P_{k_\parallel}Q_{k_\perp}
\end{equation}
of the dependent variables. Here $P_{k_\parallel}$ satisfies
\begin{equation}\label{rharm}
\hat \Delta P_{k_\parallel}=-\frac{k^2_\parallel}{a_1^2}P_{k_\parallel}\, , \quad \delta_a P_{k_\parallel}=
\dot P_{k_\parallel}=0
\end{equation}
where  $k_{\parallel}$ are the constant co-moving wave numbers in the direction of anisotropy and
the scale factor in this direction, $a_1$, is given by the zeroth order equation
\begin{equation}
\frac{\dot a_1}{a_1}=\frac{1}{3}\theta+\Sigma \, .
\end{equation}
If a coordinate $z$ is adopted to the 1-direction a possible choice for $P_{k_\parallel}$ is 
$P_{k_\parallel}=e^{ik_\parallel z}$ as
can be seen by direct substitution into (\ref{rharm}), using a tetrad, e.g. (\ref{tetrad}), adopted to the symmetries.
Similarly the harmonics $Q_{k_\perp}$ are introduced on the 2-sheets as was done in \cite{Clarkson}. They satisfy
\begin{equation}\label{qharm}
\delta^2 Q_{k_\perp}=-\frac{k^2_{\perp}}{a_2^2}Q_{k_\perp}\, , \quad \hat Q_{k_\perp}=\dot Q_{k_\perp}=0
\end{equation}
where the scale factor on the 2-sheets, $a_2$, is obtained from
\begin{equation}
\frac{\dot a_2}{a_2}=\frac{1}{3}\theta-\frac{1}{2}\Sigma \, ,
\end{equation}
and $k_{\perp}$ are the co-moving wave numbers in the perpendicular directions.

In the irrotational case there is a simple relation between the coefficients of the modes for the
hat and slash scalars.
We note that to first order $\hatcal D$ and $\slacal D$ can be written as
\begin{equation}
\hatcal D=\frac{a}{\mu}\hat\Delta\mu \quad \hbox{and} \quad \slacal D=\frac{a}{\mu}\delta^2\mu
\end{equation}
respectively, with similar expressions for the other scalars. From the commutation relations in \cite{Clarkson}
it follows that the operators $\hat\Delta$ and $\delta^2$ commute to first order when the vorticity is zero.
Hence 
\begin{equation}
\delta^2{\hatcal D}= \delta^2\left(\frac{a}{\mu}\hat\Delta\mu\right)=\hat \Delta\left(\frac{a}{\mu}\delta^2\mu\right)
=\hat\Delta{\slacal D}
\end{equation}
holds to first order.
Using the harmonic decomposition (\ref{decomp}) and equations (\ref{rharm})
and (\ref{qharm}), one has 
\begin{equation}
\fl \delta^2{\hatcal D}=
-\sum\limits_{k_{\parallel}, k_{\perp}}{\hatcal D}_{k_{\parallel} k_{\perp}}P_{k_\parallel}
\frac{k^2_{\perp}}{a_2^2} Q_{k_\perp}=
-\sum\limits_{k_{\parallel}, k_{\perp}}{\slacal D}_{k_{\parallel} k_{\perp}}
\frac{k^2_{\parallel}}{a_1^2} P_{k_\parallel} Q_{k_\perp}=\hat\Delta{\slacal D}
\end{equation}
with similar expressions for the other scalars. Hence we find the following relations
\begin{equation}\label{relhatslash}
\sla{\Psi}_{k_{\parallel} k_{\perp}}=\left(\frac{k_{\perp}}{k_{\parallel}}\right)^2 \left(\frac{a_1}{a_2}\right)^2
{\hat{\Psi}_{k_{\parallel} k_{\perp}}}\, ,
\end{equation}
where $\hat\Psi$ is $\hatcal D$, $\hatcal Z$, ${\tilde{\hatcal T}}$ or $\hatcal S$ and
$\sla\Psi$ is $\slacal D$, $\slacal Z$, ${\tilde{\slacal T}}$ or $\slacal S$,
between the coefficients of the different modes. In the following we will often suppress the subscripts
$k_{\parallel}k_{\perp}$ when it is obvious that we are refering to harmonic coefficients.

%%%%%%%%%%%%%%%%%%%%%%%%%%%%%%%%%%%%%%%%%%%%%%%%
\section{Perturbations of analytical solutions}\label{sec:analytical}
%%%%%%%%%%%%%%%%%%%%%%%%%%%%%%%%%%%%%%%%%%%%%%%%
In this section we summarize some results on perturbations around a few of the exact vacuum solutions from section 
\ref{sec:vacuum}
in the infinite wavelength ($k=0$) limit. More details are given in appendix \ref{appD}.
It might seem unphysical to consider density perturbations in a vacuum solution, but the solutions
might approximate perturbations around some background metrics with $p\ll\mu\ll\Lambda$. Furthermore,
the analytical results are useful for a comparision with results of the numerical codes. This also gives
a method of identifying suitable initial conditions.
%%%%%%%%%%%%%%%%%%%%%%%%%%%%%%%%%%%%%%%%%%%%%%%%
\subsubsection{Perturbations of $_+X$}
%%%%%%%%%%%%%%%%%%%%%%%%%%%%%%%%%%%%%%%%%%%%%%%%
With the $_+X$-solution as background 
one finds the following fourth order equation for the density gradients $\hat{\cal D}$ in the direction of anisotropy 
\begin{equation}\label{4thDX}
\hat{\cal D}^{(4)}+\frac{23}{3}\sqrt{\Lambda}\hat{\cal D}^{(3)}+\frac{59}{3}\Lambda{\ddot{\hat{\cal D}}}+
\frac{545}{27}\Lambda^{3/2}{\dot{\hat{\cal D}}}
+\frac{550}{81}\Lambda^2\hat{\cal D}=0
\end{equation}
if the harmonic numbers are put to zero (corresponding to large wavelenghts of the perturbations).
The solution ${\cal D}_{\parallel}$, i.e. $\hatcal D$ multiplied with the scale factor $a_1 
\propto  e^{\sqrt{\Lambda}t}$, is given by
\begin{equation}\label{4thDXsol}
{\cal D}_{\parallel}=A_1e^{-8\sqrt{\Lambda}t/3}+A_2e^{\sqrt{\Lambda}t/3}+(A_3 t + A_4)e^{-2\sqrt{\Lambda}t/3},
\end{equation}
where $A_1$, $A_2$, $A_3$ and $A_4$ are integration constants (or slowly varying functions over space if the
wave number is not exactly zero). Hence there is
one growing and three decaying modes.

The solution for the density gradients ${\cal D}_{\perp}=a_2 \slacal D$ in the perpendicular directions is 
then obtained from 
${\cal D}_{\perp}=a_2\left(\frac{a_1}{a_2}\right)^2\hatcal D \propto
e^{2\sqrt{\Lambda t}}\hatcal D$ as
\begin{equation}
{\cal D}_{\perp}=B_1e^{-5\sqrt{\Lambda}t/3}+B_2e^{4\sqrt{\Lambda}t/3}+(B_3 t + B_4)e^{\sqrt{\Lambda}t/3}.
\end{equation}
As seen, in these directions that are not expanding, the modes are growing faster.
%%%%%%%%%%%%%%%%%%%%%%%%%%%%%%%%%%%%%%%%%%%%%%%%
\subsubsection{Perturbations of vacuum bounce solution}
%%%%%%%%%%%%%%%%%%%%%%%%%%%%%%%%%%%%%%%%%%%%%%%%
With the vacuum bounce solution (\ref{metricbounce}) as background  
it is suitable to change the independent variable to $\theta$ through
\begin{displaymath}
\dot \psi=\frac{d \psi}{d \theta}\dot \theta =\frac{d \psi}{d \theta}(\Lambda-\theta^2) \, .
\end{displaymath}
The following equation for the density gradient $\hatcal D$ in the direction of anisotropy
\begin{eqnarray}\nonumber
\fl &&\frac{d^4 \hatcal D}{d \theta^4}+\frac{(7\theta^2+6\Lambda)}{3\theta (\theta^2-\Lambda)}\frac{d^3 \hatcal D}{d \theta^3}-
\frac{(7\theta^4-6\Lambda\theta^2-6\Lambda^2)}{3\theta^2(\theta^2-\Lambda)^2}\frac{d^2 \hatcal D}{d \theta^2}
+\frac{5(8\theta^4-45\Lambda\theta^2+18\Lambda^2)}{27\theta(\theta^2-\Lambda)^3}\frac{d \hatcal D}{d \theta}\\
\fl &&-\frac{(-735\Lambda\theta^4+495\Lambda^2\theta^2-40\theta^6-270\Lambda^3)}{81\theta^2(\theta^2-\Lambda)^4}
\hatcal D=0\label{eqDhatbounce}
\end{eqnarray}
is then obtained in the long wavelength limit. 
The solution ${\cal D}_{\parallel}=a_1\hatcal D$, where $a_1 \propto  1/\sqrt{\Lambda-\theta^2}$, is then given by
\begin{eqnarray}\nonumber
\fl {\cal D}_{\parallel}
 &=&\left(A_1+A_2\theta\right)\left(\Lambda-\theta^2\right)^{1/3}+A_3\theta\left(\Lambda-\theta^2\right)^{-1/6}
+A_4\left(\Lambda-\theta^2\right)^{1/3}\times \\ \fl &&
\left[\frac{1}{2}\ln\left(1-\frac{\theta^2}{\Lambda}\right) - \right. 
\left. \frac{\theta}{4\sqrt{\Lambda}}\ln\left(\frac{\sqrt{\Lambda}+\theta}{\sqrt{\Lambda}-\theta}\right)
 \frac{\theta}{\sqrt{\Lambda-\theta^2}}\arcsin\left(\frac{\theta}{\sqrt{\Lambda}}\right)
\right]
\label{Dhatbounce}
\end{eqnarray}
(assuming $\theta^2<\Lambda$). 
Here $\theta=\sqrt{\Lambda}\tanh(\sqrt{\Lambda}t)$ and hence $\Lambda-\theta^2=\Lambda/\cosh^2(\sqrt{\Lambda}t)$. 
The mode $A_1$ starts growing, obtains its largest value at the bounce and then starts decaying. The $A_2$
mode also starts growing, but reaches its maximum before the bounce. It then decays to zero magnitude
at the bounce and after this passes through a new maximum before it eventually decays. The $A_3$ and
$A_4$ modes initially decay,
pass through zero at $t=0$ and then grow unboundedly.

As in the previous case the growth of the density perturbations in the non-expanding directions are obtained 
from ${\cal D}_{\perp}=a_2\slacal D=a_2\left(\frac{a_1}{a_2}\right)^2\hatcal D\propto{\hatcal D}/(\Lambda-\theta^2)$ as
\begin{eqnarray}\nonumber
\fl \cal D_{\perp}&=&\left(B_1+B_2\theta\right)\left(\Lambda-\theta^2\right)^{-1/6}+
B_3\theta\left(\Lambda-\theta^2\right)^{-2/3}+
B_4\left(\Lambda-\theta^2\right)^{-1/6}\times \\ \fl &&
\left[\frac{1}{2}\ln\left(1-\frac{\theta^2}{\Lambda}\right) - \right. 
 \left. \frac{\theta}{4\sqrt{\Lambda}}\ln\left(\frac{\sqrt{\Lambda}+\theta}{\sqrt{\Lambda}-\theta}\right)
+\frac{\theta}{\sqrt{\Lambda-\theta^2}}\arcsin\left(\frac{\theta}{\sqrt{\Lambda}}\right)
\right) \, .
\label{Dslashbounce}
\end{eqnarray}
These modes decay before the bounce and grow after. The modes $B_3$ and $B_4$ are close to zero for
a longer period of time around the bounce.

For this solution perturbations with co-moving wave number $k_{\parallel}>2 a_1(t_0)\sqrt{\Lambda}$ cross
the horizon in the anisotropy direction within a finite time, and one would expect a difference 
in behaviour between perturbations with wave numbers that are smaller or larger than $2 a_1(t_0)\sqrt{\Lambda}$,
respectively. 
This is also favored by the numerical analysis in section \ref{sec:numerical}, where
typically the perturbations in similar models show an oscillatory behaviour for wave numbers
larger than this value.
%%%%%%%%%%%%%%%%%%%%%%%%%%%%%%%%%%%%%%%%%%%%%%%%
\section{Numerical solutions}\label{sec:numerical}
%%%%%%%%%%%%%%%%%%%%%%%%%%%%%%%%%%%%%%%%%%%%%%%%
The time evolution  of the density gradients are solved for numerically in some 
representative cases. To find suitable backgrounds we have used the 
phase space analysis in \cite{GoliathEllis} (see also section \ref{dynamical}).
Source points are given by the expanding Kasner metrics $_+K_\pm$
and contracting de Sitter $_-dS$, and sink points by the contracting Kasner metrics $_-K_\pm$
and expanding de Sitter $_+dS$. The plots shown below either start from expanding Kasner
$_+K_\pm$, or orginates from them. They mainly end at expanding de Sitter  $_+dS$, being
a more likely state of the late universe, but we also give examples of metrics starting from
$_+K_-$ and ending in $_-K_-$.
Saddle points are given by the flat Friedmann models $_\pm F$
and the solutions $_\pm X$, which were useful in finding solutions undergoing a bounce.
This is related to that the vacuum bounce solution (\ref{metricbounce}) starts in $_-X$ and ends at $_+X$. 
For each type of path in phase space one case with radiation and one with dust are given.

The system contains four different modes, $\cal D$, $\cal Z$, $\cal T$ and $\cal S$,
for each background and wave number.
In most of the presented cases just an initial density perturbation is assumed,
but in the two first cases we also show the effects of an initial pure shear perturbation.  
The behaviour for different types of initial perturbations has been checked, but the differences 
one can see are of the same types as those given between figures \ref{figdenspert1}
and  \ref{figshearpert1}, and figures \ref{figdenspert6} and \ref{figshearpert3} respectively.
The differences are largest in the dust case, where the density perturbations depend more
strongly on wave number with initial shear and expansion perturbations than with an
intial density perturbation due to that factors
like $\hat\Delta\left(\tilde{\hatcal T}+\tilde{\slacal T}\right)$ appear as source terms
in the equations.
For each radiation case we show the evolution for the wave numbers $k=k_{\parallel}/a_{10}=k_{\perp}/a_{20}=0,1,5$ and 20, including both sub horizontal
and super horizontal perturbations. For dust, where the dependence on wave number is
small for intial pure density perturbations, we only show the evoultion for one wave number
except for the case with intial shear perturbations.

Throughout this section we choose the cosmological constant $\Lambda=1$. 
Other initial values are given in the figure captions
or text. The code was tested against the results of section \ref{sec:analytical}.
 
%%%%%%%%%%%%%%%%%%%%%%%%%%%%%%%%%%%%%%%%%%%%%%%%
\subsection{Solutions with a bounce}
%%%%%%%%%%%%%%%%%%%%%%%%%%%%%%%%%%%%%%%%%%%%%%%%
\subsubsection{From $_-X$ via $_+X$ to $_+dS$, radiation}\label{sec:Xrad}

\begin{figure}%[h]
\hskip 1cm
\epsfxsize=4.5in
\epsffile{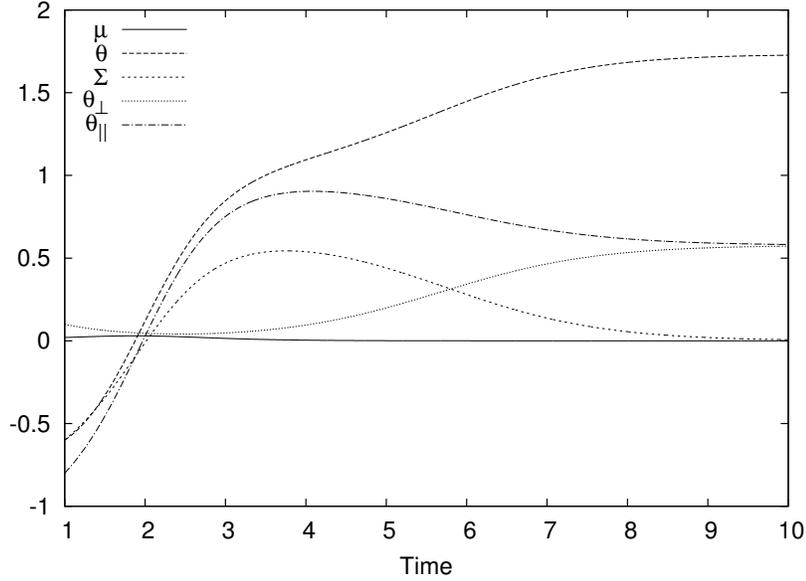} 
\caption{A radiation background that experiences a bounce. Initially it is close to $_-X$ and a late
times it approaches $_+dS$. Initial values at $t_0=1$ are given by $\mu_0=0.02$, 
$\theta_0=-0.6$, $\Sigma_0=-0.6$, $\theta_{\parallel}=-0.8$ and $\theta_{\perp}=0.1$. }
\label{figbackground1}
\end{figure}
\begin{figure}%[h]
\hskip 1cm
\epsfxsize=4.5in
\epsffile{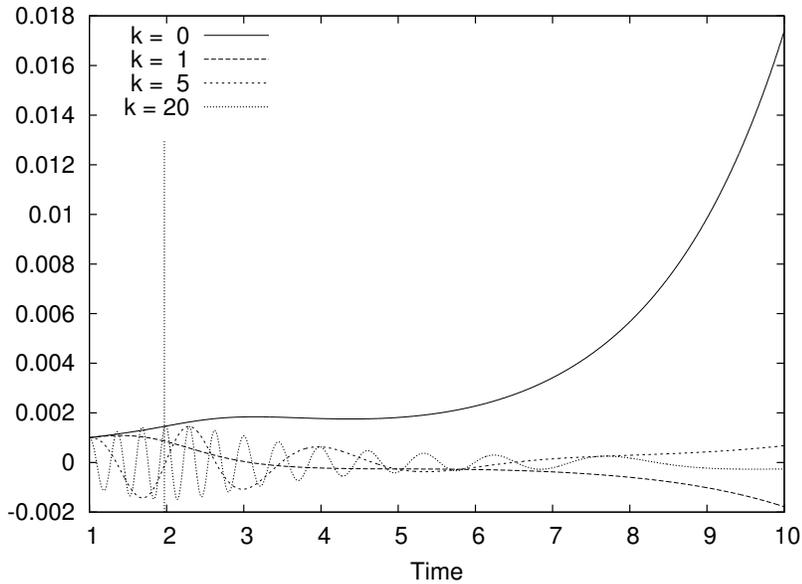} 
\caption{The growth of the density perturbations ${\cal D}_{\parallel}$ in the background given by figure \ref{figbackground1} for
the wave numbers $k=k_{\parallel}/a_{10}=k_{\perp}/a_{20}=0,1,5$ and 20.
Initially, at $t_0=1$, $\hatcal D=\slacal D=0.001$. The time of the bounce is indicated with a dotted
vertical line.
(Here $a_{10}=a_1(t_0)$
and $a_{20}=a_2(t_0)$.)}
\label{figdenspert1}
\end{figure}
\begin{figure}%[h]
\hskip 1cm
\epsfxsize=4.5in
\epsffile{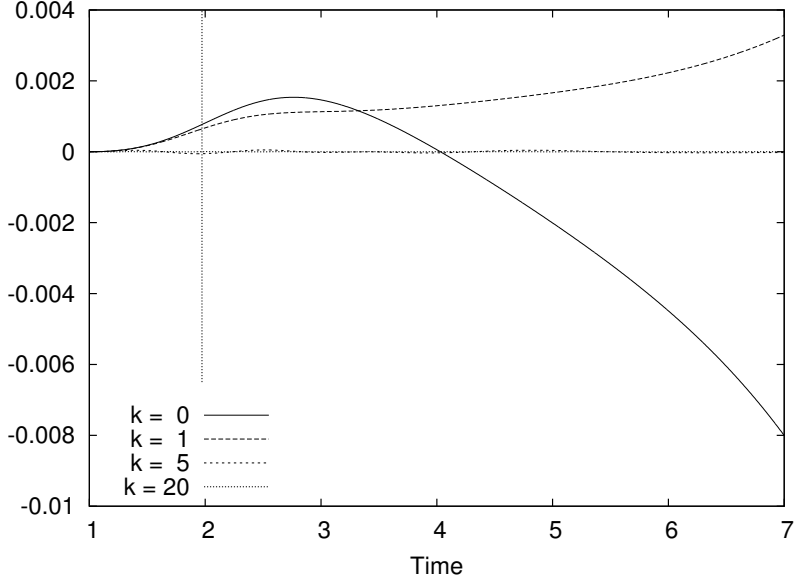} 
\caption{The growth of the density perturbations ${\cal D}_{\parallel}$ in the background given by figure \ref{figbackground1} for
the wave numbers $k=k_{\parallel}/a_{10}=k_{\perp}/a_{20}=0,1,5$ and 20. 
Initially, at $t_0=1$, ${\tilde{\hatcal T}}={\tilde{\slacal T}}=0.001$.
The time of the bounce is indicated with a dotted
vertical line.}
\label{figshearpert1}
\end{figure}

This background, where the equation of state is given by $p=\mu/3$ (radiation), passes through a state where the directions
perpendicular to the direction of anisotropy have small and
decreasing expansion rates, $\theta_{\perp}\equiv{\dot a_2}/{a_2}$, that become almost neglible for a period of time. After this the expansion
starts again to finally reach a constant value $\theta_{\perp}=\sqrt{\Lambda/3}$. The anisotropy 
direction comes from a contracting state, goes through a bounce and then starts expanding. 
Eventually the expansion rate in this direction, $\theta_{\parallel}\equiv{\dot a_1}/{a_1}$, also approaches $\theta_{\parallel}=\sqrt{\Lambda/3}$, 
giving a total expansion
rate of $\theta=\sqrt{3\Lambda}$. The initial state is close to the critical point $_-X$ and for an intermediate
period the solution is close to $_+X$, but for large
times the solution approaches the sink point de Sitter $_+dS$. The evolutions of density, $\mu_0$, 
expansion, $\theta_0$,
expansion in the anisotropy direction, $\theta_{\parallel}$, expansion in one of the perpendicular directions, $\theta_{\perp}$ 
and 
shear $\Sigma_0$ are depicted in figure \ref{figbackground1}. Note that, as is found by running time
backwards in the numerical code, that this solution can be seen as originating from the source point
$_+K_+$.

In figure \ref{figdenspert1} the growth of density perturbations for different values of the co-moving wave
numbers $k_{\parallel}$ and $k_{\perp}$ is shown. Since from equation (\ref{relhatslash}) ${\cal D}_{\perp}=
\left(\frac{k_{\perp}}{k_{\parallel}}\right)^2 \frac{a_1}{a_2}{\cal D}_{\parallel}$ we only show
${\cal D}_{\parallel}$.
The initial values of the density perturbations at $t_0=1$ are given by $\hatcal D=\slacal D=0.001$
and the other quantites are initially put to zero, ${\hatcal Z}={\slacal Z}={\tilde{\hatcal T}}={\tilde{\slacal T}}={\hatcal S}={\slacal S}=0$. 
For $k=0$ the density gradient in the direction of anisotropy reaches a small maximum after the bounce and after
this a small minimum before it starts growing unboundedly. In the perpendicular directions the gradient for the $k=0$ case 
is roughly constant before it also starts growing.
For higher values of the wave number $k$ the density gradient in the anisotropy direction shows an oscillatory behaviour with an initally
increasing amplitude that later on decreases, but does not fall of to zero.
This behaviour with a local maximum 
in the density gradient at or slightly
after the bounce in the bouncing direction seems to be typical. 
For corresponding $k$ values in the perpendicular directions the oscillations initially have an approximately constant amplitude
that with time slowly starts growing.

In \ref{appE} the corresponding growth of density perturbations in a closed and isotropic radiation filled
universe undergoing a bounce (now in all three directions) is shown in figure \ref{figPertFriedmann}. The local extrema
seen in figure \ref{figdenspert1} are absent in the isotropic case, apart from a small dip for the $k=0$ mode,
whereas the behaviours for large times, when the universes
in both cases approach de Sitter spacetime, are similar.

In figure \ref{figshearpert1} the growth of density perturbations for
a case where the density perturbations initially are zero, $\hatcal D=\slacal D=0$, is shown.
Instead there is an initial shear perturbation ${\tilde{\hatcal T}}={\tilde{\slacal T}}=0.001$
(and $\hatcal Z=\slacal Z=\hatcal S=\slacal S=0$). As seen the super horizon modes $k=0,1$
grows unboundedly, whereas for higher $k$-values the amplitudes remain small.
%%%%%%%%%%%%%%%%%%%%%%%%%%%%%%%%%%%%%%%%%%%%%%%%
\subsubsection{From $_-X$ via $_+X$ to $_+dS$, dust}\label{sec:Xdust}
%%%%%%%%%%%%%%%%%%%%%%%%%%%%%%%%%%%%%%%%%%%%%%%%
\begin{figure}%[h]
\hskip 1cm
\epsfxsize=4.5in
\epsffile{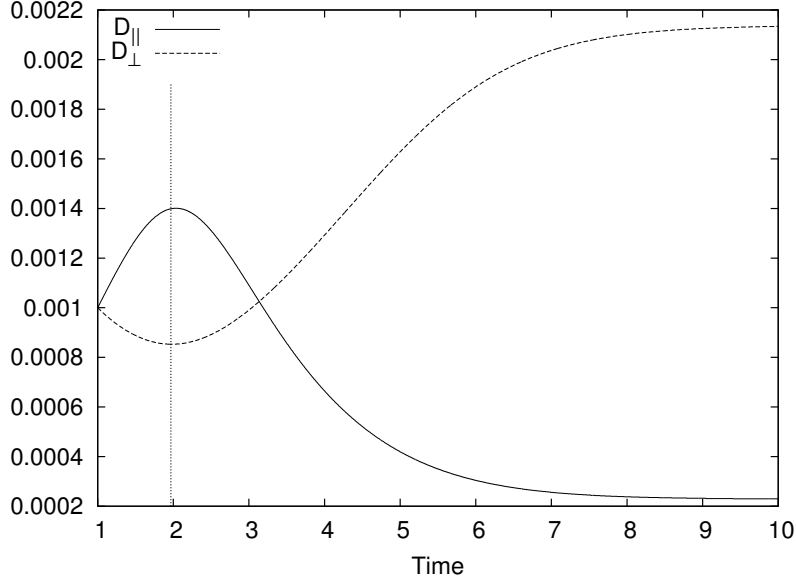} 
\caption{The growth of the density perturbations ${\cal D}_{\parallel}$ and ${\cal D}_{\perp}$
in the dust background of section \ref{sec:Xdust} for
the wave numbers $k=k_{\parallel}/a_{10}=k_{\perp}/a_{20}=1$. Initially, at $t_0=1$, $\hatcal D=\slacal D=0.001$
and $\mu_0=0.02$, 
$\theta_0=-0.6$, $\Sigma_0=-0.6$, $\theta_{\parallel}=-0.8$ and $\theta_{\perp}=0.1$.
The time of the bounce is indicated with a dotted
vertical line.}
\label{figdenspert6}
\end{figure}
\begin{figure}%[h]
\hskip 1cm
\epsfxsize=4.5in
\epsffile{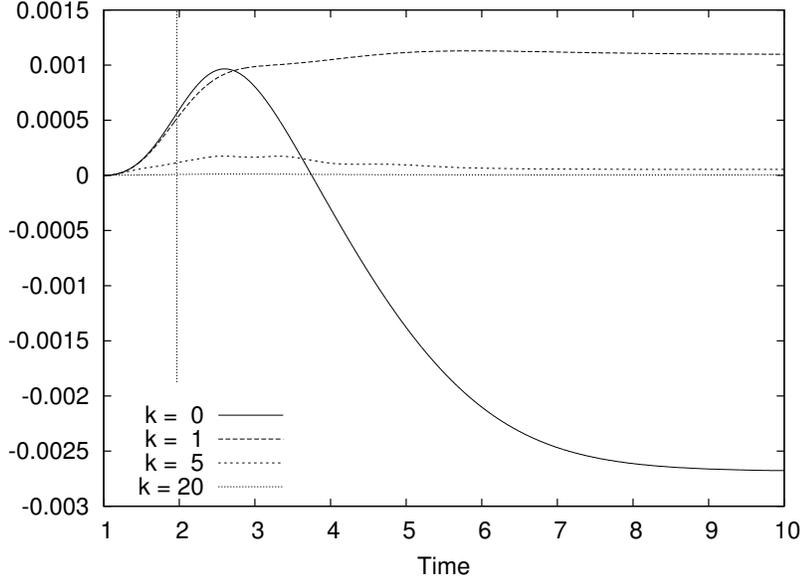} 
\caption{The growth of the density perturbations ${\cal D}_{\parallel}$ in the dust background of section 
\ref{sec:Xdust} for
the wave numbers $k=k_{\parallel}/a_{10}=k_{\perp}/a_{20}=0,1,5$ and 20. 
Initially, at $t_0=1$, ${\tilde{\hatcal T}}={\tilde{\slacal T}}=0.001$.
The time of the bounce is indicated with a dotted
vertical line.}
\label{figshearpert3}
\end{figure}

The conditions here are similar to those in the previous section, but the equation of state is now given by dust, $p=0$.
With the same initial conditions for the background quantities as in the radiation case, their evolutions are close
to those of the radiation case, see figure \ref{figbackground1}. 
First we show the time development of the density perturbations for the case when initially
$\hatcal D=\slacal D=0.001$
and $\hatcal Z=\slacal Z={\tilde{\hatcal T}}={\tilde{\slacal T}}=\hatcal S=\slacal S=0$. 
Since in this case the evolution of the density perturbations is relatively insensitive to wave number, we here only give
them for $k=1$  in figure \ref{figdenspert6}. 
In the direction of anisotropy the density gradients first grow and reaches a maximum at approximately the time of the
bounce and then decay into a small but nonzero value. For the perpendicular directions the behaviour is inversed. First
a minimum is reached around the time of the bounce, and then the gradient grows towards a constant value. 
A comparison with the evolution of density perturbations in a bouncing closed and isotropic dust universe, see figure
\ref{figPertFriedmanndust} in \ref{appE}, shows that the local minimum (maximum) is absent in the isotropic case.

Next, in picture \ref{figshearpert3}, the growth of the density 
perturbations ${\cal D}_{\parallel}$ for the case when there is an initial shear perturbation 
${\tilde{\hatcal T}}={\tilde{\slacal T}}=0.001$ (and the other perturbations are zero), is shown.
In this case the density perturbations depend on the wave number $k$. The reason
for this can be seen from the systems of equations (\ref{eqhat1})-(\ref{eqhat4}) and (\ref{eqslash1})-
(\ref{eqslash4}), where $\hat\Delta\left(\tilde{\hatcal T}+\tilde{\slacal T}\right)$ and
$\delta^2\left(\tilde{\hatcal T}+\tilde{\slacal T}\right)$ appear as source terms.

Note that all perturbations asymptotically approach constant values. This is consistent
with that the background asymptotically approaches de Sitter. The result holds
also for the cases \ref{dustFriedmanndeSitter} and \ref{dustKasnerdeSitter}, that both are
dust solutions approaching de Sitter. That this result does not apply for the radiation 
solutions can be understood from that terms like $p/\mu=1/3$ remain in the equations
also in the limit $\mu \rightarrow 0$.  
%%%%%%%%%%%%%%%%%%%%%%%%%%%%%%%%%%%%%%%%%%%%%%%%
\subsection{Recollapsing solutions}
%%%%%%%%%%%%%%%%%%%%%%%%%%%%%%%%%%%%%%%%%%%%%%%%
%%%%%%%%%%%%%%%%%%%%%%%%%%%%%%%%%%%%%%%%%%%%%%%%
\subsubsection{From $_+K_-$ to $_-K_-$, radiation}\label{sec:kasnerrad}
%%%%%%%%%%%%%%%%%%%%%%%%%%%%%%%%%%%%%%%%%%%%%%%%
\begin{figure}
\hskip 1cm
\epsfxsize=4.5in
\epsffile{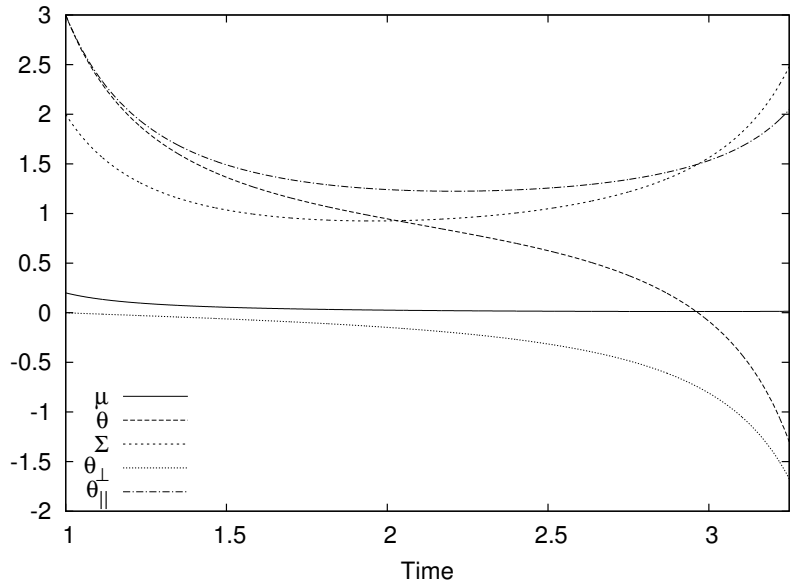} 
\caption{A radiation background that starts expanding, reaches a largest value and then recollapses. It starts close to $_+K_-$ and ends at $_-K_-$. 
Initial values at $t_0=1$ are given by $\mu_0=0.2$, $\theta_0=3$, $\Sigma_0=2$, $\theta_{\parallel}=3$ and $\theta_{\perp}=0$.}
\label{figbackground3}
\end{figure}
\begin{figure}
\hskip 1cm
\epsfxsize=4.5in
\epsffile{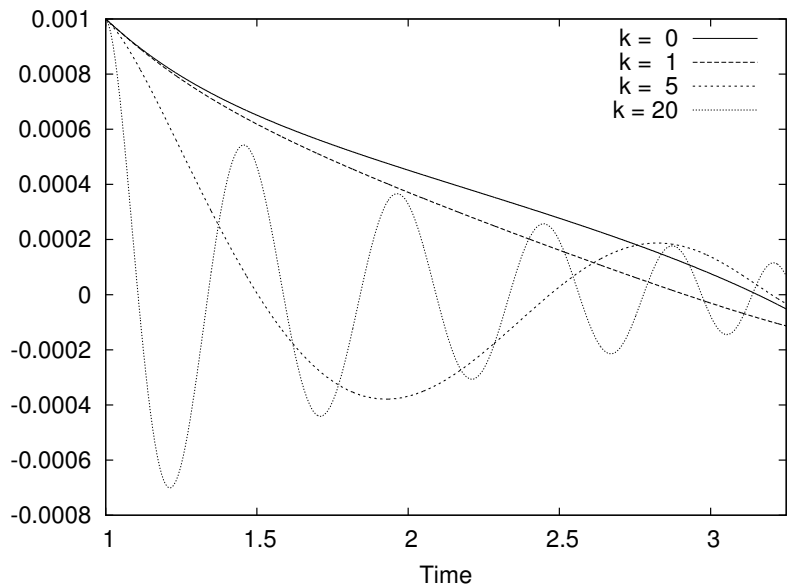} 
\caption{The growth of the density perturbations ${\cal D}_{\parallel}$ in the background given by figure \ref{figbackground3} for
the wave numbers $k=k_{\parallel}/a_{10}=k_{\perp}/a_{20}=0,1,5$ and 20. 
Initially $\hatcal D=\slacal D=0.001$.}
\label{figdenspert8}
\end{figure}

Here the background, that is filled with radiation, starts close to an expanding Kasner, $_+K_-$, reaches a largest value, 
and then approaches
a collapsing Kasner, $_-K_-$. In the direction of anisotropy it is always expanding, whereas it is collapsing in the perpendicular directions. 
Initially this collapse velocity is negligible, but eventually it will dominate over the
expansion in the direction of anisotropy. The evolutions of density, $\mu_0$, 
expansion, $\theta_0$, expansion in anisotropy direction, $\theta_{\parallel}$, expansion in one of the perpendicular directions, 
$\theta_{\perp}$ and shear $\Sigma_0$ are depicted in figure \ref{figbackground3}.
In figure \ref{figdenspert8} the growth of the density perturbations ${\cal D}_{\parallel}$ for different values of the comoving wave
numbers $k_{\parallel}$ and $k_{\perp}$ is shown. Initially $\hatcal D=\slacal D=0.001$ and
$\hatcal Z=\slacal Z={\tilde{\hatcal T}}={\tilde{\slacal T}}=\hatcal S=\slacal S=0$. 

As can be seen the density gradient in the direction of anisotropy decays for
all $k$-numbers. In the collapsing directions the gradients grow.  

%%%%%%%%%%%%%%%%%%%%%%%%%%%%%%%%%%%%%%%%%%%%%%%%
\subsubsection{From $_+K_-$ to $_-K_-$, dust}\label{sec:kasnerdust}
%%%%%%%%%%%%%%%%%%%%%%%%%%%%%%%%%%%%%%%%%%%%%%%%
\begin{figure}
\hskip 1cm
\epsfxsize=4.5in
\epsffile{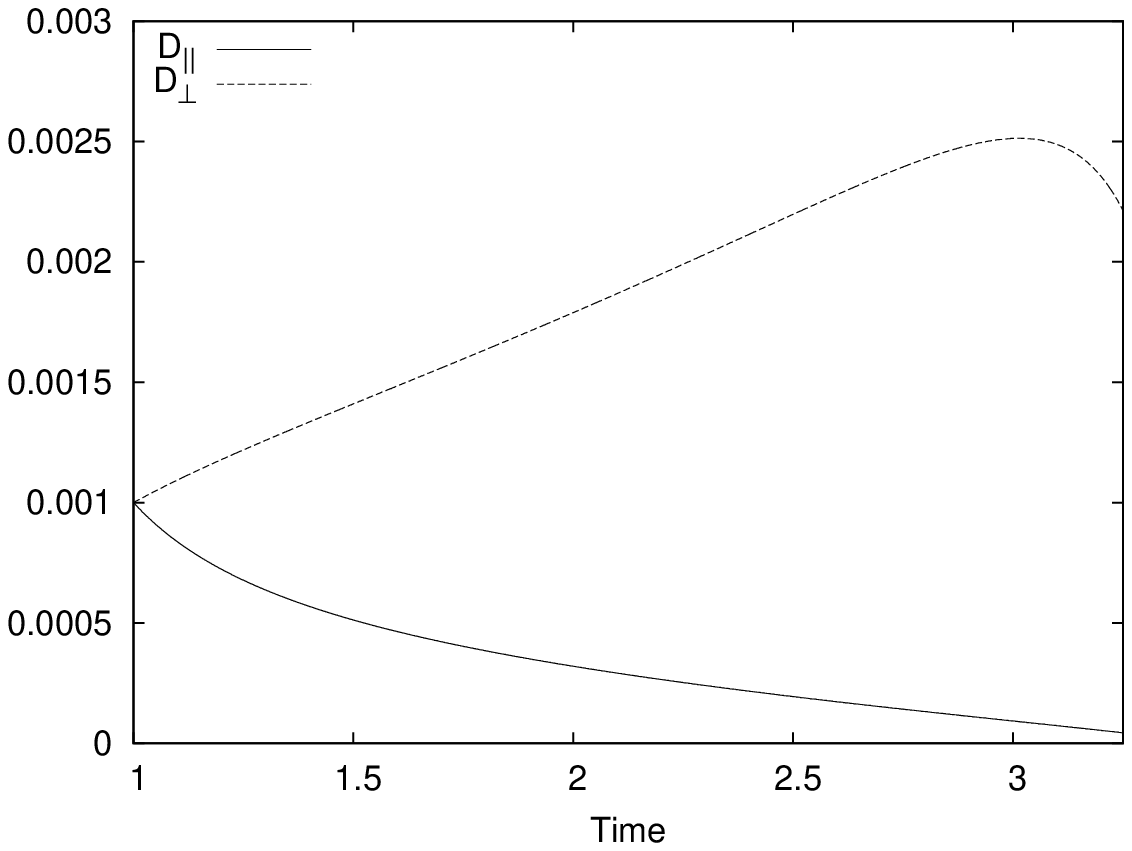} 
\caption{The growth of density perturbations ${\cal D}_{\parallel}$ and ${\cal D}_{\perp}$
in the dust background of section \ref{sec:kasnerdust} for
the wave numbers $k=k_{\parallel}/a_{10}=k_{\perp}/a_{20}=1$. 
Initially $\hatcal D=\slacal D=0.001$ and $\mu_0=0.2$, $\theta_0=3$, $\Sigma_0=2$, $\theta_{\parallel}=3$ and $\theta_{\perp}=0$.}
\label{figdenspert13}
\end{figure}

This is a similar situation to the previous one, but with dust.
The evolutions of the background quantities are once again close to those of the corresponding radiation
case in \ref{sec:kasnerrad}, see figure \ref{figbackground3}. Also in this dust
case the perturbations are
rather insensitive to the wave number, and we hence only give them for one $k$-value in figure \ref{figdenspert13}. 
As before the the density perturbations decay in the expanding direction and initially grow in the contracting directions.
%%%%%%%%%%%%%%%%%%%%%%%%%%%%%%%%%%%%%%%%%%%%%%%%
\subsection{Solutions without a bounce}
%%%%%%%%%%%%%%%%%%%%%%%%%%%%%%%%%%%%%%%%%%%%%%%%
For all cases in this section the initial conditions of the perturbations are $\hatcal D=\slacal D=0.001$ and
$\hatcal Z=\slacal Z={\tilde{\hatcal T}}={\tilde{\slacal T}}=\hatcal S=\slacal S=0$.
%%%%%%%%%%%%%%%%%%%%%%%%%%%%%%%%%%%%%%%%%%%%%%%%
\subsubsection{Friedmann to de Sitter, radiation}\label{radFriedmanndeSitter}
%%%%%%%%%%%%%%%%%%%%%%%%%%%%%%%%%%%%%%%%%%%%%%%%
\begin{figure}
\hskip 1cm
\epsfxsize=4.5in
\epsffile{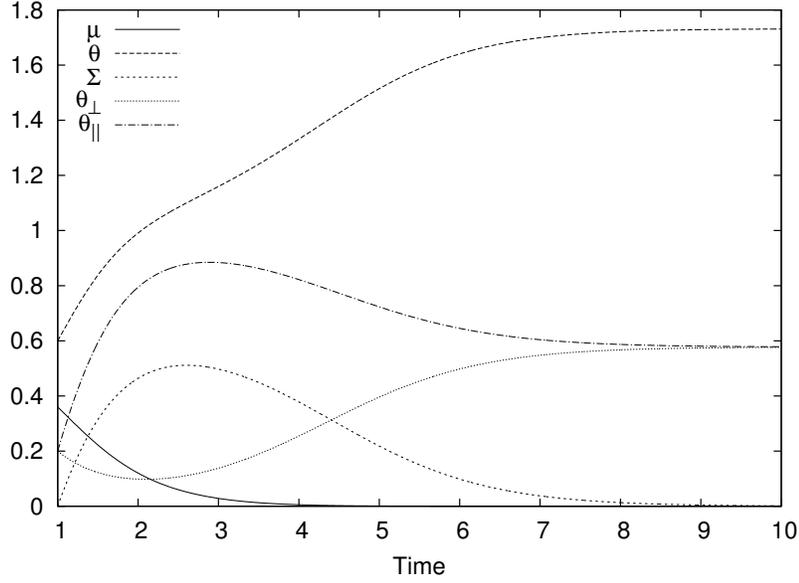} 
\caption{A radiation background passing close to expanding Friedmann, $_+F$, and ending at expanding de Sitter 
$_+dS$. Initial values at $t_0=1$ are given by $\mu_0=0.36$, $\theta_0=0.6$ and $\Sigma_0=0$.}
\label{figbackground5}
\end{figure}
\begin{figure}
\hskip 1cm
\epsfxsize=4.5in
\epsffile{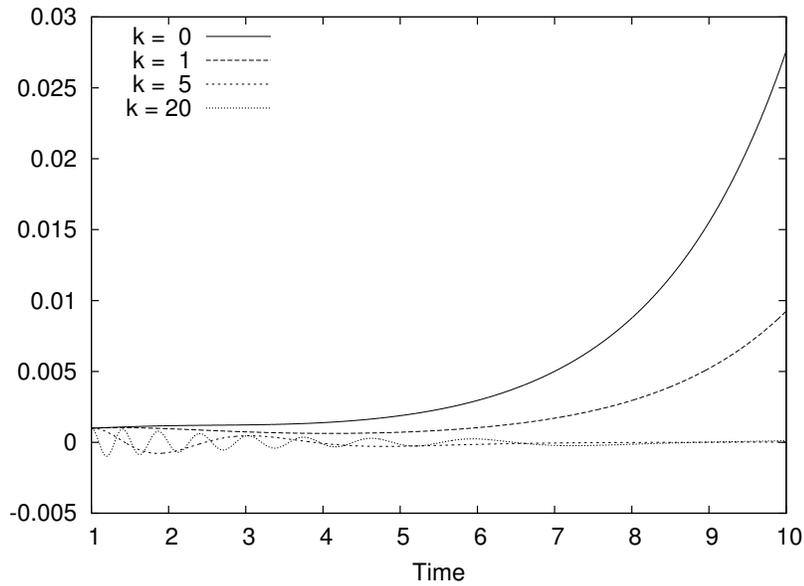} 
\caption{The growth of the density perturbations ${\cal D}_{\parallel}$ in the background given by figure \ref{figbackground5} for
the wave numbers $k=k_{\parallel}/a_{10}=k_{\perp}/a_{20}=0,1,5$ and 20.
Initially $\hatcal D=\slacal D=0.001$.}
\label{figdenspert14}
\end{figure}

This radiation background passes close to an expanding Friedmann, $_+F$, and ends at an expanding de Sitter,
$_+dS$. It originates from an expanding Kasner, $_+K_+$, as can be seen by integrating backwards
in time with the same initial coditions. Note also that at $t=1$, the starting point in figure  \ref{figbackground5}, the growth rate of the shear is nonzero.
The evolutions of density, $\mu_0$, 
expansion, $\theta_0$, expansion in anisotropy direction, $\theta_{\parallel}$, expansion in one of the perpendicular directions, 
$\theta_{\perp}$ and shear $\Sigma_0$ are depicted in figure \ref{figbackground5}. The
expansion in the direction of anisotropy is for a period of time dominating over the expansion in the
perpendicular directions.
In figure \ref{figdenspert14} the growth of the density perturbations ${\cal D}_{\parallel}$ for different values of the comoving wave
numbers $k_{\parallel}$ and $k_{\perp}$ is shown. 
In the direction of anistropy the super horizon modes $k=0,1$ eventually grows unboundedly, whereas for higher modes the amplitude
slowly falls of. In the perpendicular directions the $k=0,1$ modes grows faster and the amplitudes of the higher modes slowly grow.
%%%%%%%%%%%%%%%%%%%%%%%%%%%%%%%%%%%%%%%%%%%%%%%%
\subsubsection{Friedmann to de Sitter, dust}\label{dustFriedmanndeSitter}
%%%%%%%%%%%%%%%%%%%%%%%%%%%%%%%%%%%%%%%%%%%%%%%%
\begin{figure}
\hskip 1cm
\epsfxsize=4.5in
\epsffile{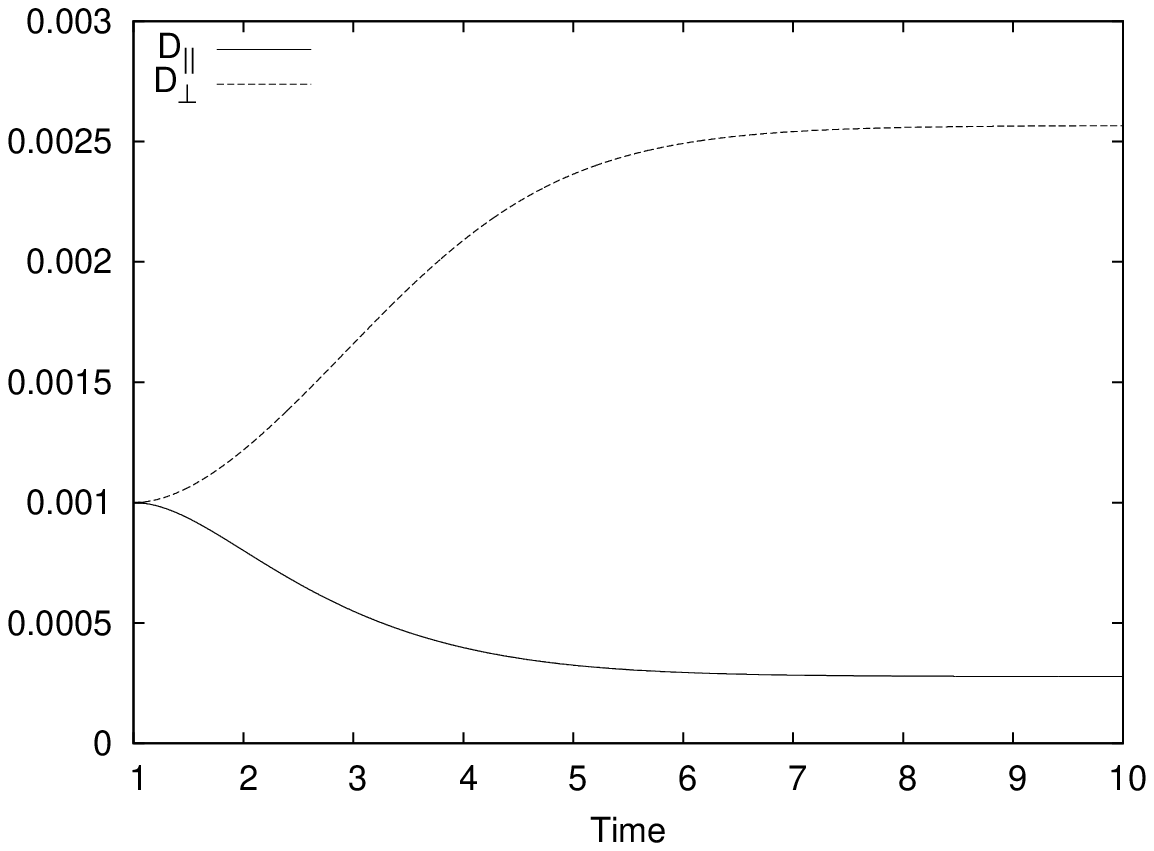} 
\caption{The growth of density perturbations ${\cal D}_{\parallel}$ and ${\cal D}_{\perp}$
in the dust background of section \ref{dustFriedmanndeSitter} for
the wave numbers $k=k_{\parallel}/a_{10}=k_{\perp}/a_{20}=1$.
Initially $\hatcal D=\slacal D=0.001$ and $\mu_0=0.36$, $\theta_0=0.6$ and $\Sigma_0=0$.}
\label{figdenspert19}
\end{figure}

For the corresponding dust background to the radiation case of section \ref{radFriedmanndeSitter},
the evolutions of the background quantities are once again close to those of the radiation case, 
see figure \ref{figbackground5}. 
As in the previous dust models the perturbations are insensitive to the wave number.
In figure \ref{figdenspert19} the growth of density perturbations for the comoving wave
number $k=1$ are shown. The density gradients in the direction of anisotropy falls of to a nonzero constant value,
whereas in the perpendicular directions they grow towards a higher constant value.
%%%%%%%%%%%%%%%%%%%%%%%%%%%%%%%%%%%%%%%%%%%%%%%%
\subsubsection{$_+K_-$ to $_+dS$, radiation}\label{radKasnerdeSitter}
%%%%%%%%%%%%%%%%%%%%%%%%%%%%%%%%%%%%%%%%%%%%%%%%
\begin{figure}
\hskip 1cm
\epsfxsize=4.5in
\epsffile{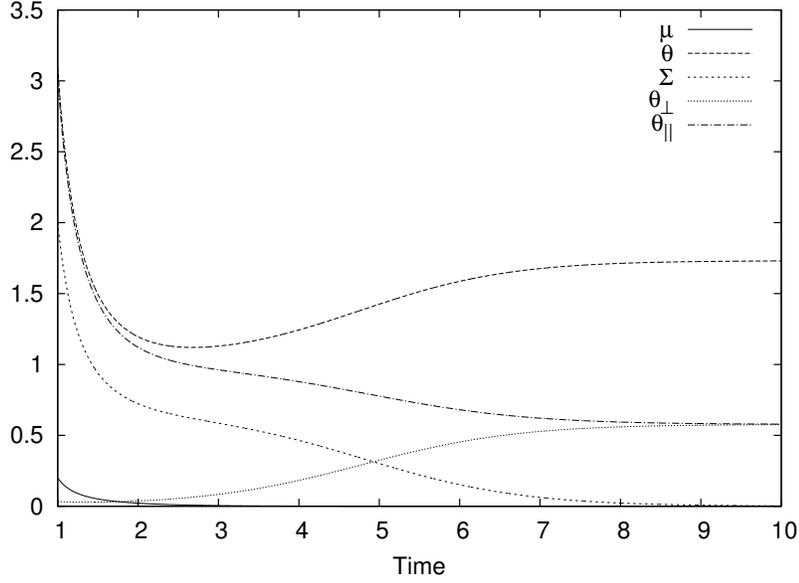} 
\caption{A radiation background starting close to expanding Kasner, $_+K_-$, and ending at expanding de Sitter 
$_+dS$. Initial values at $t_0=1$ are given by $\mu_0=0.2$, $\theta_0=3.1$ and $\Sigma=2$.}
\label{figbackground7}
\end{figure}
\begin{figure}
\hskip 1cm
\epsfxsize=4.5in
\epsffile{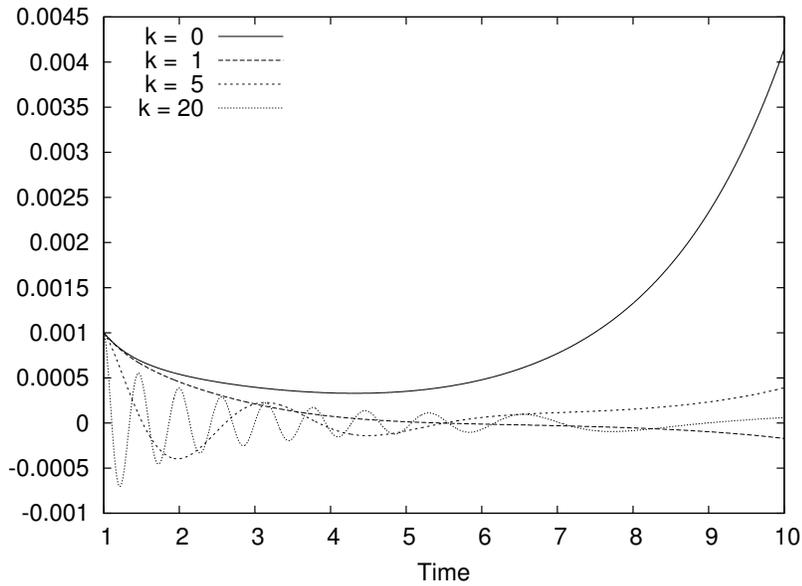} 
\caption{The growth of the density perturbations ${\cal D}_{\parallel}$ in the background given by figure \ref{figbackground7} for
the wave numbers $k=k_{\parallel}/a_{10}=k_{\perp}/a_{20}=0,1,5$ and 20.
Initially $\hatcal D=\slacal D=0.001$.}
\label{figdenspert21}
\end{figure}

This radiation background starts close to an expanding Kasner, $_+K_-$, and ends at an expanding de Sitter,
$_+dS$. The evolutions of density, $\mu_0$, 
expansion, $\theta_0$, expansion in anisotropy direction, $\theta_{\parallel}$, expansion in one of the perpendicular directions, 
$\theta_{\perp}$ and shear $\Sigma_0$ are depicted in figure \ref{figbackground7}. 
In figure \ref{figdenspert21} the growth of the density perturbations ${\cal D}_{\parallel}$ for different values of the comoving wave
numbers $k_{\parallel}$ and $k_{\perp}$ is shown. 
In the direction of anisotropy the $k=0$ mode initially decreases, but then turns and starts to grow unboundedly. The higher modes initially have
decreasing amplitudes. Also these starts to increase, but more slowly, at about the same time as the $k=0$ mode. In the perpendicular directions the 
modes do not show this initial decrease, but otherwise
behave in a similar way.
%%%%%%%%%%%%%%%%%%%%%%%%%%%%%%%%%%%%%%%%%%%%%%%%
\subsubsection{$_+K_-$ to $_+dS$, dust}\label{dustKasnerdeSitter}
%%%%%%%%%%%%%%%%%%%%%%%%%%%%%%%%%%%%%%%%%%%%%%%%
\begin{figure}
\hskip 1cm
\epsfxsize=4.5in
\epsffile{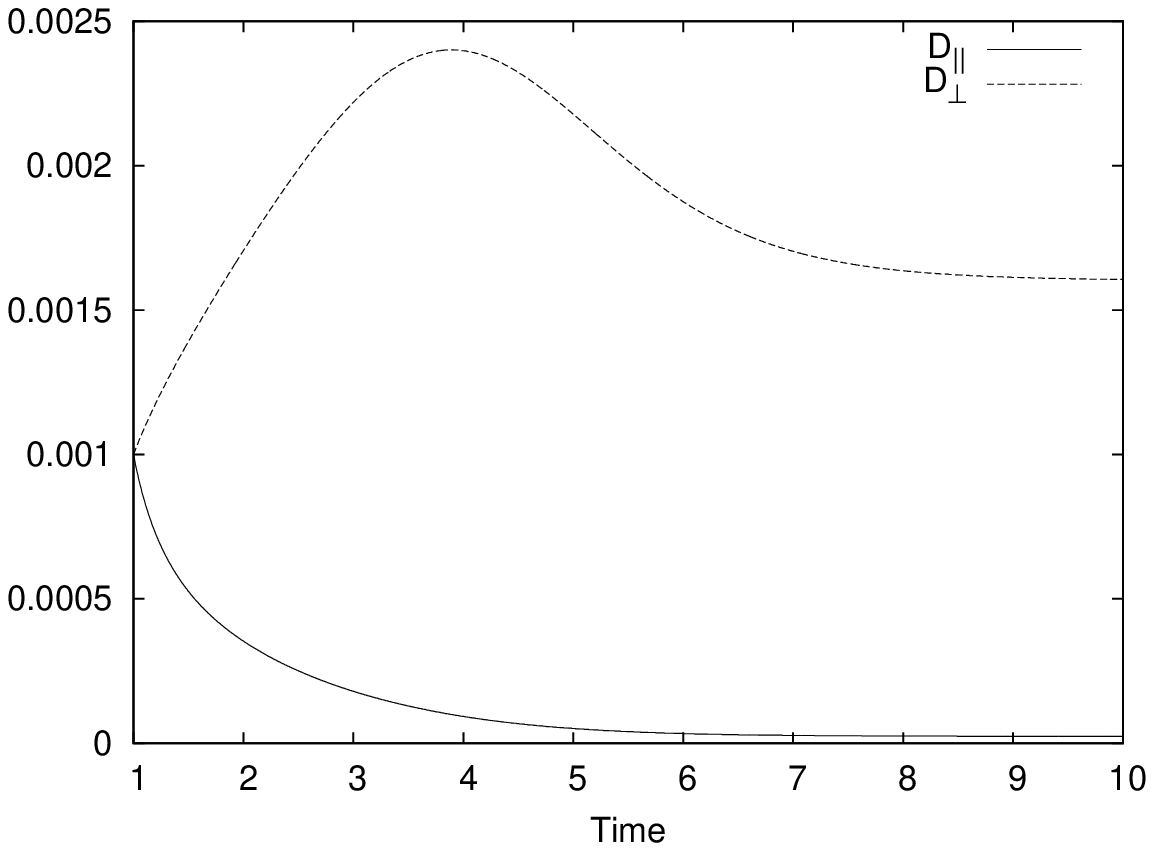} 
\caption{The growth of density perturbations ${\cal D}_{\parallel}$ and ${\cal D}_{\perp}$
in the dust background in section \ref{dustKasnerdeSitter} for
the wave numbers $k=k_{\parallel}/a_{10}=k_{\perp}/a_{20}=1$. 
Initially $\hatcal D=\slacal D=0.001$ and $\mu_0=0.2$, $\theta_0=3.1$ and $\Sigma=2$.}
\label{figdenspert25}
\end{figure}

This dust background starts close to an expanding Kasner, $_+K_-$, and ends at an expanding de Sitter,
$_+dS$. For the evolutions of the background quantities see figure \ref{figbackground7} for the corresponding
radiation case, which shows a similar behaviour.
In figure \ref{figdenspert25} the growth of density perturbations for $k=1$ is given. 
%%%%%%%%%%%%%%%%%%%%%%%%%%%%%%%%%%%%%%%%%%%%%%%%
\section{Summary}
%%%%%%%%%%%%%%%%%%%%%%%%%%%%%%%%%%%%%%%%%%%%%%%%
A closed system for scalar perturbations on Kantowski-Sachs cosmologies has been found in terms
of gauge invariant variables. For long wavelengths and low densities some
analytical results are obtained. By redefining some apparently singular terms, the system could be
rewritten in a form suitable for numerical analysis. Due to the complexity of the governing equations, 
the choice of background, initial conditions and wave numbers, many different behaviors of the
growth of the density perturbations can be obtained. In general the growth of density gradients
is different in the direction of the anisotropy and in the perpendicular directions, which should
influence the formation of structures. 

More importantly,  growth rates are
also affected by a bounce, see, e.g., figure \ref{figdenspert6}. Typically the density gradients 
experience maxima (or minima) slightly after the bounce in the anisotropic Kantowski-Sachs models.
A comparison with density growths in bouncing closed FLRW models, see \ref{appE}, 
shows that the amplitudes of the perturbations are roughly constant initially and that the
local extrema around the time of the bounce are absent or very small in the isotropic case.

It was shown in \cite{Ehlers} that a FLRW model with dust, radiation and cosmological constant
cannot undergo a bounce, unless it took place after the formation of the quasars. In section
\ref{sectionbounce} we use similar arguments in the Kantowski-Sachs case to show that a bounce in the directions perpendicular
to the preferred direction cannot take place at all, whereas a bounce in the preferred direction cannot 
be excluded by these arguments. We also derive relations between present values
of observational quantities including expansion and deceleration rates in the different directions.
Since anistropies in the Hubble and decelerations parameters cannot be excluded by present observations,
see e.g. \cite{McClureDyer} and \cite{CaiTuo}, it would be of interest to get better estimates
of these parameters.

The present analysis can be extended to study tensor perturbations, 
including gravitational waves, and the coupling between scalar and tensor perturbations. In this way 
possible observational consequences of anisotropic bounces can be formulated in a way which makes it easy to
compare with what happens in the standard case. The techniques developed here can also be used to develop a general framework 
for describing the evolution of cosmological perturbations in other non--standard cosmologies for example the Lema\^{i}tre--Tolman--Bondi models \cite{Uzan}. Such studies are extremely important as they test the various rigidities which exist between different set of independent observables ( from the background and at the level of perturbation theory) that can be used to test the underlying hypothesis of the standard cosmological model.

%%%%%%%%%%%%%%%%%%%%%%%%%%%%%%%%%%%%%%%%%%%%%%%%
\ack%nowledgments
%%%%%%%%%%%%%%%%%%%%%%%%%%%%%%%%%%%%%%%%%%%%%%%%
This work was supported by SIDA (Sweden) and NRF (South Africa) grants. ZK was supported by the OTKA grant
no. 100216.
MB and MF gratefully acknowledges the hospitality of the department of Mathematics and Applied Mathematics at UCT.
We thank the referees for useful suggestions which helped to improve the paper.
%%%%%%%%%%%%%%%%%%%%%%%%%%%%%%%%%%%%%%%%%%%%%%%%
\appendix
\section{Some approximate background-solutions with matter}\label{appA}
%%%%%%%%%%%%%%%%%%%%%%%%%%%%%%%%%%%%%%%%%%%%%%%%
Here some approximate solutions with 
$\mu\sim p \ll\Lambda$ to the system (\ref{approx1})-(\ref{approx3}) are given. 

With the bounce solution (\ref{metricbounce}) as background the solution is given by
\begin{eqnarray}
\mu_1&=& C\cosh^{-\gamma}(\sqrt{\Lambda}t)\\\label{integral1}
\Sigma_1-\frac{2}{3}\theta_1&=&\cosh(\sqrt{\Lambda}t)\times 
\left[D+
\gamma C\int \frac{dt}{\cosh^{\gamma+1}(\sqrt{\Lambda} t)}\right]\\\label{integral2}
\Sigma_1+\frac{5}{6}\theta_1&=&\cosh^{-2}(\sqrt{\Lambda}t)\times \left[E + \right. 
\left.
\frac{(6-5\gamma)C}{4}\int\cosh^{2-\gamma}(\sqrt{\Lambda}t)dt\right]
\end{eqnarray}
where $C$, $D$ and $E$ are constants of integration.
The integrals in (\ref{integral1}) and (\ref{integral2}) can be performed for dust, $\gamma=1$, giving
\begin{eqnarray}
\frac{C}{\sqrt{\Lambda}}\tanh(\sqrt{\Lambda}t) \quad \hbox{and} \quad
\frac{C}{4\sqrt{\Lambda}}\sinh(\sqrt{\Lambda}t)
\end{eqnarray}
respectively, and also for stiff matter, $\gamma=2$, giving
\begin{equation}
\frac{2C}{\sqrt{\Lambda}}\left(\frac{\sinh(\sqrt{\Lambda}t)}{2\cosh^2(\sqrt{\Lambda}t)}+
\arctan\left(e^{\sqrt{\Lambda}t}\right)\right) \quad \hbox{and} \quad \frac{Ct}{4}
\end{equation}
respectively.

With the critical points $\!_{\pm}X$, (\ref{Xmetrics}), as background, the solutions are
given by
\begin{eqnarray}
\mu_1&=&Ce^{\mp\gamma\sqrt{\Lambda}t}\\
\Sigma_1-\frac{2}{3}\theta_1&=&De^{\pm\sqrt{\Lambda}t}
\mp\frac{\gamma C}{(\gamma+1)\sqrt{\Lambda}}e^{\mp\gamma\sqrt{\Lambda}t}\\
\Sigma_1+\frac{5}{6}\theta_1&=&Ee^{\mp2\sqrt{\Lambda}t}
\pm\frac{(6-5\gamma)C}{4(2-\gamma)\sqrt{\Lambda}}e^{\mp\gamma\sqrt{\Lambda}t} \, .
\end{eqnarray}
Note that since for all of these solutions some terms grow unboundedly, they are only valid for limited
time intervals. 
%%%%%%%%%%%%%%%%%%%%%%%%%%%%%%%%%%%%%%%%%%%%%%%%
\section{Calculation of some first order quantities}\label{appB}
%%%%%%%%%%%%%%%%%%%%%%%%%%%%%%%%%%%%%%%%%%%%%%%%
In the calculations several contractions of the type $x_{ab}y^{ab}$ of PSTF tensors
\begin{equation}
x_{ab}=X(n_a n_b-\frac{1}{2}N_{ab})+2X_{(a}n_{b)}+X_{ab}
\end{equation}
will appear. To first order contractions of two, three or four PSTF tensor are given by
\begin{eqnarray}
\fl x_{ab}y^{ab}&=&\frac{3}{2}XY \, , \quad x_{ab}y^{bc}z_c\!^a=\frac{3}{4}XYZ \, , \quad
x_{ab}y^{bc}z_{cd}u^{da}=\frac{9}{8}XYZU
\end{eqnarray}
respectively, so that, e.g., $\sigma^2 \equiv \frac{1}{2}\sigma_{ab}\sigma^{ab}=\frac{3}{4}\Sigma^2$.

In calculating the expression for $\dot S\equiv (\sigma^{ab}S_{ab})\dot{}\;$, needed for the
equation for $\dot {\cal S}_a$, the following term
\begin{eqnarray}\nonumber
\fl &&\sigma^{ab}({\rm{curl}} H)_{ab}=\sigma^{ab}({\rm{curl}}({\rm{curl}} \sigma))_{ab}\equiv 
\sigma^{ab}\eta^{cd}\!_{<a}\tilde\nabla_c({\rm{curl}}\sigma)_{b>d} = \\
\fl &&\sigma^{ab}\eta^{cd}\!_{a}\tilde\nabla_c({\rm{curl}}\sigma)_{bd}\equiv
\sigma^{ab}\eta^{cd}\!_{a}\tilde\nabla_c\eta^{ef}\!_{<b}\tilde\nabla_e\sigma_{d>f}
=\sigma^a\!_b\eta^{c}\!_{da}\eta^{ef<b}\tilde\nabla_c\tilde\nabla_e\sigma^{d>}\!_{f} \, ,
\end{eqnarray}
where it was used that $\sigma_{ab}$ already is PSTF and that the projected derivative of
$\eta_{abc}$ is zero, will appear. Using the definition of PSTF and that
\begin{equation}
\eta_{abc}\eta^{def}= \left| \begin{array}{ccc}
     h_a^d     &  h_a^e   &  h_a^f  \\ 
     h_b^d    &   h_b^e  &  h_b^f  \\ 
     h_c^d   &   h_c^e  &   h_c^f \\ 
\end{array}
\right|
\end{equation}
it simplifies to 
\begin{eqnarray}\nonumber
&&\sigma^{ab}({\rm{curl}}H)_{ab}=
-\tilde\nabla_a\tilde\nabla^a(\sigma^2)+\frac{1}{2}\sigma^{ab}\tilde\nabla_b\tilde\nabla^c\sigma_{ca}+ 
\sigma_{ab}\tilde\nabla_c\tilde\nabla^a\sigma^{bc} \\  &&
=-\tilde\nabla_a\tilde\nabla^a(\sigma^2)+
\frac{1}{2}\sigma^{ab}\tilde\nabla_b(\frac{2}{3}\tilde\nabla_a\theta) 
+\sigma_{ab}\tilde\nabla_c\tilde\nabla^a\sigma^{bc} \, ,
\end{eqnarray}
where (\ref{divsigma}) was used in the last step. With the definitions (\ref{defZ}) and
the 3-dimensional Ricci identity
\begin{equation}
\tilde\nabla_c\tilde\nabla_d\sigma_{ab}-\tilde\nabla_d\tilde\nabla_c\sigma_{ab}=
^{(3)}\!\!R_{aecd}\sigma^e_b+^{(3)}\!\!R_{ebcd}\sigma^e_a \; ,
\end{equation}
that holds with $\tilde\nabla_a$ since $\omega_{ab}=0$, one gets
\begin{eqnarray}\label{eqcurlcurl}
\fl \sigma^{ab}({\rm{curl}}H)_{ab}=-\frac{1}{a}\tilde\nabla_a{\cal T}^a+
\frac{1}{a}\sigma^{ab}\tilde\nabla_a{\cal Z}_b+
\sigma^{ab}\,  ^{(3)}\!R_{acdb}\sigma^{cd}+
^{(3)}\!\!R_{ab}\sigma^{bc}\sigma_c^a \, .
\end{eqnarray}
The 3-Riemann tensor is given in terms of the projected 4-curvature, $R_{abcd\perp}\equiv h_a^eh_b^fh_c^gh_d^h R_{efgh}$, 
and the extrinsic curvature, $K_{ab}\equiv\tilde\nabla_{(a}u_{b)}
=\theta/3+\sigma_{ab}$, as
\begin{equation}
^{(3)}\!R_{abcd}=R_{abcd\perp}-K_{ac}K_{bd}+K_{bc}K_{ad}
\end{equation}
and the 3-Ricci tensor by
\begin{eqnarray}
^{(3)}\!R_{ab}&=&S_{ab}+\frac{1}{3}h_{ab}\;^{(3)}\!R= 
S_{ab}+\frac{1}{3}h_{ab}\left(2\mu-\frac{2}{3}\theta+2\sigma^2+2\Lambda\right) \, .
\end{eqnarray}
Then, on using that the 4-curvature can be decomposed as
\begin{eqnarray}\nonumber
R^{ab}\!\!_{cd}&=&\frac{2}{3}(\mu +3p-2\Lambda)u^{[a}u_{[c}h^{b]}\!_{d]}+\frac{2}{3}h^{a}\!_{[c}h^b\!_{d]}+\\
&&4u^{[a}u_{[c}E^{b]}\!_{d]}+4h^{[a}\!_{[c}E^{b]}\!_{d]}+2\eta^{abe}u_{[c}H_{d]e}+ 
2\eta_{cde}u^{[a}H^{b]e}
\end{eqnarray}
(\ref{eqcurlcurl}) becomes
\begin{eqnarray}
\fl \sigma^{ab}({\rm{curl}}H)_{ab}&=&
\frac{1}{a}\sigma^{ab}\tilde\nabla_a{\cal Z}_b-\frac{1}{a}\tilde\nabla_a{\cal T}^a+2\left(\sigma^2\right)^2-
\frac{2}{3}\theta^2\sigma^2
+3\tilde S\sigma^2+2\sigma^2(\mu+\Lambda)
\end{eqnarray}
where $\tilde S$ is defined from equation (\ref{Stilde}).
%%%%%%%%%%%%%%%%%%%%%%%%%%%%%%%%%%%%%%%%%%%%%%%%
\section{Perturbative equations in orthogonal directions}\label{appC}
%%%%%%%%%%%%%%%%%%%%%%%%%%%%%%%%%%%%%%%%%%%%%%%%
The perturbative equations, obtained by projecting (\ref{eqD})-(\ref{eqS}) with $N_{ab}$, are
\begin{eqnarray}\label{dv}
\fl \dot {\cal D}_{\bar a}& =&({\cal D}_{\bar a})^.=\frac{\theta p}{\mu}{\cal D}_{\bar a}+
\frac{1}{2}\Sigma{\cal D}_{\bar a}-{\cal Z}_{\bar a}(1+\frac{p}{\mu}) \\\nonumber
\fl \dot {\cal Z}_{\bar a} &=&({\cal Z}_{\bar a})^.=-\frac{2}{3}\theta {\cal Z}_{\bar a}-
\frac{1}{2}\mu{\cal D}_{\bar a}+\frac{1}{2}\Sigma {\cal Z}_{\bar a}-2{\cal T}_{\bar a}+  
\frac{3}{2}\frac{\mu p'}{\mu+p}\left(\tilde S+\frac{3}{2}\Sigma^2\right){\cal D}_{\bar a} \\ \label{zv}
\fl &&-\frac{\mu p'}{\mu+p}\delta_a\tilde\nabla^b{\cal D}_b \\\nonumber
\fl \dot {\cal T}_{\bar a} &=&({\cal T}_{\bar a})^.= -2\theta{\cal T}_{\bar a}+\frac{1}{2}\Sigma{\cal T}_{\bar a}-
{\cal S}_{\bar a}-\frac{3}{2}\Sigma^2{\cal Z}_{\bar a}+ 
\frac{3}{2}\Sigma\left(\tilde S + \theta\Sigma\right)
\frac{\mu p'}{\mu + p}{\cal D}_{\bar a}- \\ \label{tv}
\fl &&-\frac{\mu p'}{\mu + p}\Sigma(\frac{3}{2}  \delta_an^c \tilde\nabla_c{\cal D}-
\frac{1}{2}\delta_a\tilde\nabla^b{\cal D}_b) 
\\\nonumber
\fl \dot {\cal S}_{\bar a}&=&({\cal S}_{\bar a})^.=\left(\Sigma^2+2\frac{\tilde S^2}{\Sigma^2}\right){\cal T}_{\bar a}+
\left(\frac{5}{2}\Sigma-\frac{5}{3}\theta-
2\frac{\tilde S}{\Sigma}\right){\cal S}_{\bar a}+
\frac{p'\mu}{\mu+p}\tilde S\left(\frac{5}{2}\theta\Sigma+\frac{3}{2}{\tilde S}-\frac{3}{2}\Sigma^2
\right){\cal D}_{\bar a}\\\nonumber 
\fl &&+\mu\Sigma^2 {\cal D}_{\bar a}-\Sigma\left(\frac{5}{2}\tilde S +
\frac{2}{3}\Sigma\theta\right){\cal Z}_{\bar a}+
\frac{p'\mu}{\mu+p}\left[\frac{1}{2}\left({\tilde S}-
\frac{1}{3}\theta\Sigma+2\Sigma^2\right)\delta_a\tilde\nabla^b{\cal D}_b - \right. \\\label{sv}
\fl &&\left. \frac{3}{2}\left(\tilde S-\frac{1}{3}\theta\Sigma+\Sigma^2\right) \delta_a n^b\tilde\nabla_b{\cal D}\right]
+\frac{3}{2}\Sigma  \delta_a n^b\tilde\nabla_b{\cal Z}-
\frac{1}{2}\Sigma\delta_a\tilde\nabla^b{\cal Z}_b-\delta_a\tilde\nabla^b{\cal T}_b \, .
\end{eqnarray}
The system obtained by taking the 2-divergence of (\ref{dv})-(\ref{sv}) is
\begin{eqnarray}\label{eqslash1}
\fl \dot {\slacal D}  &=& \bras{ \theta \bra{\frac{ p}{\mu} - \frac{1}{3}} +\Sigma } \slacal D - \bra{1+\frac{p}{\mu}} 
\slacal Z \\ \nonumber
\fl \dot {\slacal Z}  &=& \bra{  \Sigma-\theta  } \slacal Z  +
\bras{ -\frac{1}{2}\mu + \frac{3}{2}\frac{\mu p'}{\mu + p}\bra{\tilde S + \frac{3}{2}\Sigma^2} } \slacal D 
 - 2 \Sigma^2\tilde{\slacal T}-2\frac{\Sigma}{\tilde S}\slacal S \\
\fl && - \frac{\mu p'}{\mu + p} \delta^2 \bras{  \hat{\cal D} +   \slacal{D}} 
\end{eqnarray}
\begin{eqnarray}\nonumber
\fl \dot {\tilde{\slacal T}}  &=& -\bra{\frac{1}{3}\theta - \Sigma+\frac{\Sigma^3}{\tilde S}}\tilde{ \slacal T} - 
\left(\frac{\Sigma^2}{\tilde S^2}+\frac{1}{\tilde S}\right)\slacal S - 
\left[\frac{\Sigma \mu}{\tilde S}+\frac{\mu p'}{\mu + p} \left(\theta-\frac{3}{2}\Sigma\right)\right]\slacal D 
+\left(1+ \frac{2}{3}\frac{ \Sigma\theta}{\tilde S}\right) \slacal Z
 + 
\\ \nonumber
\fl &&
\frac{\mu p'}{\mu + p}\frac{1}{\tilde S} \bras{\left(\frac{1}{2}\Sigma-\frac{1}{3}\theta\right) \delta^2 \hatcal D -
\left(\Sigma- \frac{1}{6}\theta\right) \delta^2 \bra{\slacal{D}} } 
-\frac{1}{\tilde S}\delta^2(\hatcal Z-\frac{1}{2}\slacal Z)
+\frac{\Sigma}{\tilde S}\delta^2({\tilde{\hatcal T}}+{\tilde{\slacal T}})+
\\ 
\fl &&\frac{1}{\tilde S^2} \delta^2
(\hatcal S+ \slacal S)
\\ \nonumber
\fl \dot {\slacal S} &=& \bras{ \mu \Sigma^2 +
\frac{\mu p'}{\mu + p} \tilde S \bra{ \frac{5}{2}\theta \Sigma + \frac{3}{2} \tilde S - 
\frac{3}{2} \Sigma^2 } } \slacal D -
\bra{\frac{2}{3} \theta \Sigma + \frac{5}{2} \tilde S } \Sigma \slacal Z  
  + \bra{\Sigma^4 + 2 \tilde S^2} \tilde{\slacal T} + 
\\ \nonumber
\fl &&\bra{\frac{\Sigma^3}{\tilde S}-2\theta +3\Sigma } \slacal S
+ \Sigma  \delta^2\left( \hatcal Z - \frac{1}{2}  \slacal{Z}\right) 
	- \Sigma^2 \delta^2 \bra{{\tilde{\hatcal T}}+{\tilde{\slacal{T}}}} 
-\frac{\Sigma}{\tilde S}\delta^2(\hatcal S+\slacal S)+
  \\ 
\fl &&	\frac{\mu p'}{\mu + p}\left[\bra{-\tilde S + \frac{1}{3} \theta \Sigma - \frac{1}{2} \Sigma^2}  
\delta^2 \hatcal D+ 
\right.  
 \left. 
\frac{1}{2}\bra{\tilde S - \frac{1}{3}\theta \Sigma + 2 \Sigma^2  } \delta^2 \bra{
\slacal{D}}\right]\, .
\label{eqslash4}
\end{eqnarray}
%%%%%%%%%%%%%%%%%%%%%%%%%%%%%%%%%%%%%%%%%%%%%%%%
\section{Perturbations around analytical solutions}\label{appD}
%%%%%%%%%%%%%%%%%%%%%%%%%%%%%%%%%%%%%%%%%%%%%%%%
\subsection{Perturbations around $_+X$}
%%%%%%%%%%%%%%%%%%%%%%%%%%%%%%%%%%%%%%%%%%%%%%%%
With the $_+X$-solution as background the system (\ref{eqhat1})-(\ref{eqhat4}) for the hat-variables reduces to

\begin{eqnarray}
\dot {\hat{\cal D} } &=& -\frac{5}{3}\sqrt{\Lambda} \hat{\cal D} -  \hat{\cal Z}\, , \quad 
\dot {\hat{\cal Z} } =-  \frac{7}{3}\sqrt{\Lambda}\hat{\cal Z} - 2 \hat{\cal T} \\
\dot {\hat{\cal T}}  &=& -  \frac{11}{3}\sqrt{\Lambda}\hatcal T - \hatcal S - \frac{2}{3} \Lambda \hatcal Z \, , \quad 
\dot {\hatcal S} = \frac{22}{27}\Lambda^{3/2}\hatcal Z+\frac{22}{9}\Lambda\hatcal T 
\end{eqnarray}
if the harmonic numbers are put to zero (corresponding to large wave-lenghts of the perturbations).
From these equations one finds the fourth order equation (\ref{4thDX}) for the density perturbations $\hat{\cal D}$
with solution (\ref{4thDXsol}).
The other quantities are then obtained as
\begin{eqnarray}
\hat {\cal Z}&=&-{\dot {\hat{\cal D}}}-\frac{5}{3}\sqrt{\Lambda}{\hat{\cal D}}\, , \quad
\hat {\cal T}=\frac{1}{2}{\ddot {\hat {\cal D}}}+2\sqrt{\Lambda}+\frac{35}{18}\Lambda{\hat {\cal D}}\\
\hat {\cal S}&=&-\frac{1}{2}{\hat {\cal D}}^{(3)}-\frac{23}{6}\sqrt{\Lambda}{\ddot {\hat {\cal D}}}-\frac{155}{18}\Lambda{\dot {\hat{\cal D}}}-
\frac{325}{54}\Lambda^{3/2}\hat {\cal D}
\end{eqnarray}
giving 
\begin{eqnarray}
\hat {\cal D}&=&A_1+A_2+A_4 \, , \quad
\hat {\cal Z}=\sqrt{\Lambda}(2A_1-A_2)-A_3\\
\hat {\cal T}&=&\frac{\Lambda}{3}(4A_1+\frac{5}{2}A_2)+\frac{\sqrt{\Lambda}}{3}A_3 \, , \quad
\hat {\cal S}=-\frac{\Lambda^{3/2}}{3}(4A_1+\frac{11}{2}A_2)
\end{eqnarray}
as intial conditions at $t=0$.
%%%%%%%%%%%%%%%%%%%%%%%%%%%%%%%%%%%%%%%%%%%%%%%%
\subsection{Perturbations around vacuum bounce solution}
%%%%%%%%%%%%%%%%%%%%%%%%%%%%%%%%%%%%%%%%%%%%%%%%
With the vacuum bounce solution (\ref{metricbounce}) as background the system (\ref{eqhat1})-(\ref{eqhat4}) for the hat-variables reduces to
\begin{eqnarray}\label{eqdothatDbounce}
\dot {\hat{\cal D} } &=& -\frac{5}{3}\theta \hat{\cal D} -  \hat{\cal Z} \\ 
\dot {\hat{\cal Z} } &=&-  \frac{7}{3}\theta\hat{\cal Z} - \frac{8}{9}\theta^2 {\tilde{ \hat{\cal T}}}+
2\frac{\theta}{\Lambda}\hatcal S \\
\dot {\tilde{\hat{\cal T}}}  &=& -  \frac{1}{3}\theta\left(5-\frac{4}{3}\frac{\theta^2}{\Lambda}\right){\tilde{\hatcal T}}
 +\frac{3}{2\Lambda}\left(1-\frac{2}{3}\frac{\theta^2}{\Lambda}\right) \hatcal S
+ 
\left(1- \frac{2}{3} \frac{\theta^2}{\Lambda}\right) \hatcal Z \\ 
\dot {\hatcal S} &=& \frac{2}{9}\theta\left(5\Lambda-\frac{4}{3}\theta^2\right)\hatcal Z+
\frac{8}{9}\left(\Lambda^2+\frac{2}{9}\theta^4\right){\tilde{\hatcal T}}- 
2\theta\left(1+\frac{2}{9}\frac{\theta^2}{\Lambda}\right)\hatcal S  \label{eqdothatSbounce}
\end{eqnarray}
(once again only considering the long wave-length limit). It is suitable to change the independent variable to $\theta$ through
\begin{displaymath}
\dot \psi=\frac{d \psi}{d \theta}\dot \theta =\frac{d \psi}{d \theta}(\Lambda-\theta^2) \, .
\end{displaymath}
The equation (\ref{eqDhatbounce}) for $\hatcal D$
with solution (\ref{Dhatbounce})
is then obtained. 
The other quantities are given by
\begin{eqnarray} 
\hatcal Z &=&  - {\displaystyle \frac {5}{3}} \,\theta \,{\hatcal D} - 
(\Lambda  - 
\theta ^{2}){\frac {d \hatcal D}{d \theta }}
\end{eqnarray}
\begin{eqnarray}\nonumber
\tilde{\hatcal T} &=& \frac{1}{8\theta^2\Lambda}\left(\frac{5}{3}\left(9\Lambda^2-45\Lambda\theta^2-8\theta^4\right)
{\hatcal D}-\right. 
5\theta\left(3\Lambda+4\theta^2\right)\left(\Lambda- 
\theta^2\right)\frac {d {\hatcal D}}{d \theta } + \\ 
&& \left. 3\left(3\Lambda-5\theta^2\right)\left(\Lambda-\theta^2\right)\frac {d ^{2}{\hatcal D}}{d \theta ^{2}}
-9\theta\left(\Lambda-\theta^2\right)^3\frac{d^3 {\hatcal D}}{d\theta^3}\right)
\,\\\nonumber
\hatcal S &=& {\displaystyle -\frac {5}{54}} \,{\hatcal D}\,\theta \,
(57\,\Lambda  + 8\,\theta ^{2}) - 
{\displaystyle \frac {1}{18}} 
\,{\frac {d \hatcal D}{d \theta }}\,\,(33\,\Lambda
  + 20\,\theta ^{2})\,(\Lambda  - \theta ^{2}) - \\
&&{\displaystyle 
\frac {5}{6}} \,{\frac {d ^{2}\hatcal D}{d \theta ^{2}}}\,
\,\theta \,(\Lambda  - \theta ^{2})^{2} 
- {\displaystyle \frac {1}{2}} \,{\frac {d
^{3}\hatcal D}{d \theta ^{3}}}\,\,(\Lambda  - \theta ^{2})^{3}\, .
\end{eqnarray}
Taylor expanding $\hatcal D$ to second order around $\theta=0$ gives
\begin{eqnarray}
\hatcal D&=& \tilde A_1+(\tilde A_2+\tilde A_3)\theta-\frac{5}{6}\frac{\tilde A_1}{\Lambda}\theta^2+...\\
\hatcal Z&=&-(\tilde A_2+\tilde A_3)\Lambda+(\frac{11}{6}\tilde A_2+\frac{1}{3}\tilde A_3)\theta^2+...\\
\tilde{\hatcal T} &=&  -\frac{27}{8}\tilde A_4-\frac{9}{4}\tilde A_3\theta+\frac{9}{16}\frac{\tilde A_4}{\Lambda}\theta^2 +...\\
\hatcal S &=& (\frac{2}{3}\tilde A_2-\frac{5}{6}\tilde A_3)\Lambda^2-3\tilde A_4\Lambda\theta- 
(\frac{11}{9}\tilde A_2+\frac{13}{18}\tilde A_3)
\Lambda\theta^2+...,
\end{eqnarray}
where $\tilde A_1=A_1\Lambda^{5/6}$, $\tilde A_2=A_2\Lambda^{5/6}$, $\tilde A_3=A_3\Lambda^{1/3}$ and $\tilde A_4=A_4\Lambda^{5/6}$. 
%%%%%%%%%%%%%%%%%%%%%%%%%%%%%%%%%%%%%%%%%%%%%%%%
\section{Perturbations on bouncing closed Friedmann}\label{appE}
\begin{figure}
\hskip 1cm
\epsfxsize=4.5in
\epsffile{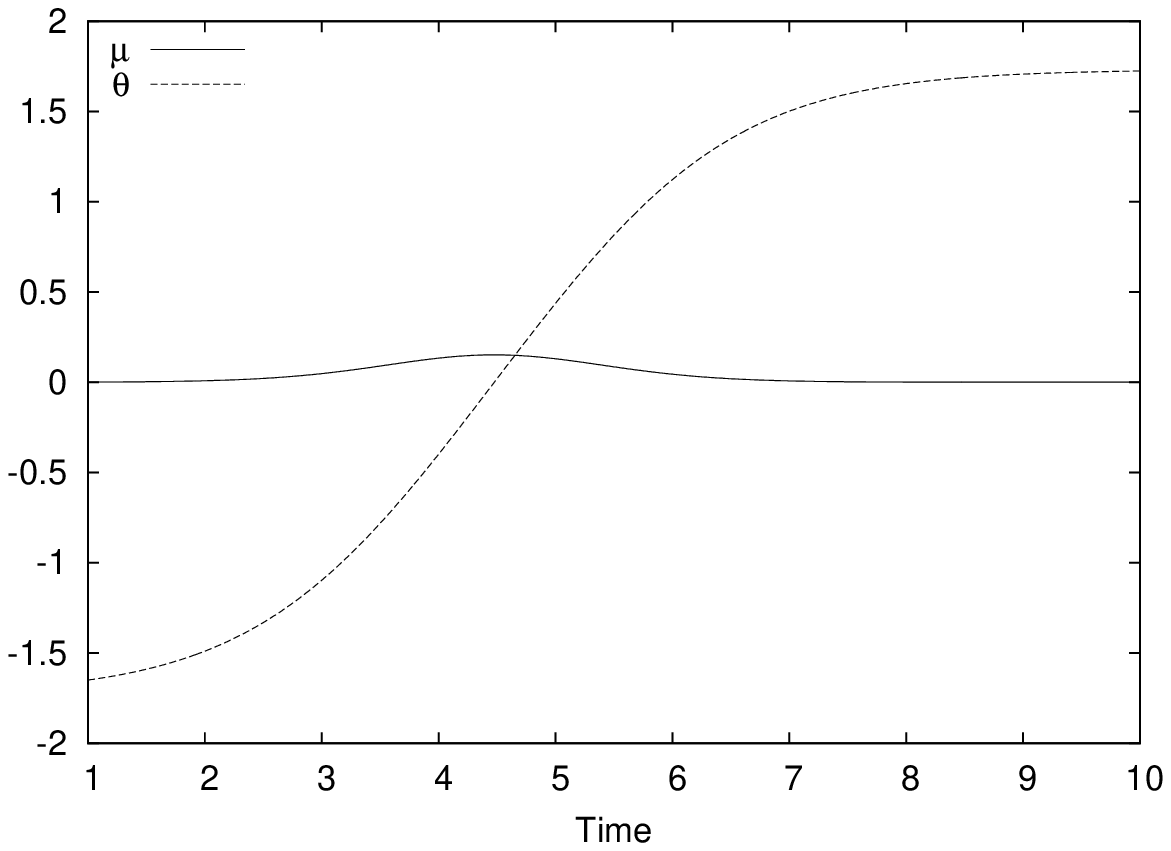} 
\caption{A closed initially contracting Friedmann model with radiation, that goes through a bounce and then starts expanding.
Initial values at $t_0=1$ are given by $\mu_0=0.001$, $\theta_0=-1.65$. $\Lambda=1$ and $C=2.96$.}
\label{figFriedmann}
\end{figure}

Closed Friedmann models with a positive cosmological constant may undergo a bounce \cite{GoliathEllis}. The
evolution equations are given by
\begin{eqnarray}\label{eqclosedmu}
\dot \mu&=&-\theta (\mu + p) \\\label{eqclosedtheta}
\dot \theta &=& -\frac{1}{3}\theta^2-\frac{1}{2}(\mu+3p-2\Lambda) \, .
\end{eqnarray}
With an equation of state $p=(\gamma-1)\mu$ the following integral
\begin{equation}\label{eqclosedintegral}
\theta^2=3\mu+3\Lambda-3C\mu^{2/3\gamma}
\end{equation}
is found. Here closed models correspond to $C>0$. In figure \ref{figFriedmann} a bouncing radiation solution is shown.
It starts from a contracting state, goes through a bounce and then asymptotically approaches de Sitter. 

To study the growth of density perturbations it is sufficient to use the density gradient
${\cal D}_a = \frac{a\tilde \nabla_a \mu}{\mu}$ and expansion gradient
${\cal Z}_a = a\tilde \nabla_a\theta$, in which the system now closes \cite{EllisvanElst}:
\begin{eqnarray}
\fl \dot {\cal D}_a =\frac{\theta p}{\mu}{\cal D}_a-{\cal Z}_a(1+\frac{p}{\mu}) \\ \label{eqDFriedmann}
\fl \dot {\cal Z}_a =-\frac{2}{3}\theta {\cal Z}_a-\frac{1}{2}\mu{\cal D}_a+ 
\frac{\mu p'}{\mu+p}\left(-\mu-\Lambda+\frac{1}{3}\theta^2\right){\cal D}_a-
\frac{\mu p'}{\mu+p}\tilde\nabla_a\tilde\nabla^b{\cal D}_b \, .
\end{eqnarray}
Here $p'\equiv dp/d\mu$.
In accordance with equations (\ref{newvar}) and (\ref{DpDp}) we define the scalar density gradient as
\begin{equation}
{\cal D}=a\tilde\nabla^a{\cal D}_a \, .
\end{equation}
A harmonic decomposition
\begin{equation}
{\cal D}=\sum\limits_{\tilde{k}}{\cal D}_{\tilde{k}}Q_{\tilde{k}} \quad \hbox{where} \quad
\tilde\nabla^2 Q_{\tilde{k}}=-\frac{\tilde k^2}{a^2} Q_{\tilde{k}} \quad \hbox{and} \quad \dot Q_{\tilde{k}}=0 
\end{equation}
is then done. Here $\tilde k^2=n(n+2)$ for $n=1,2,...$, \cite{Harrison,Tsagas}.

\begin{figure}
\hskip 1cm
\epsfxsize=4.5in
\epsffile{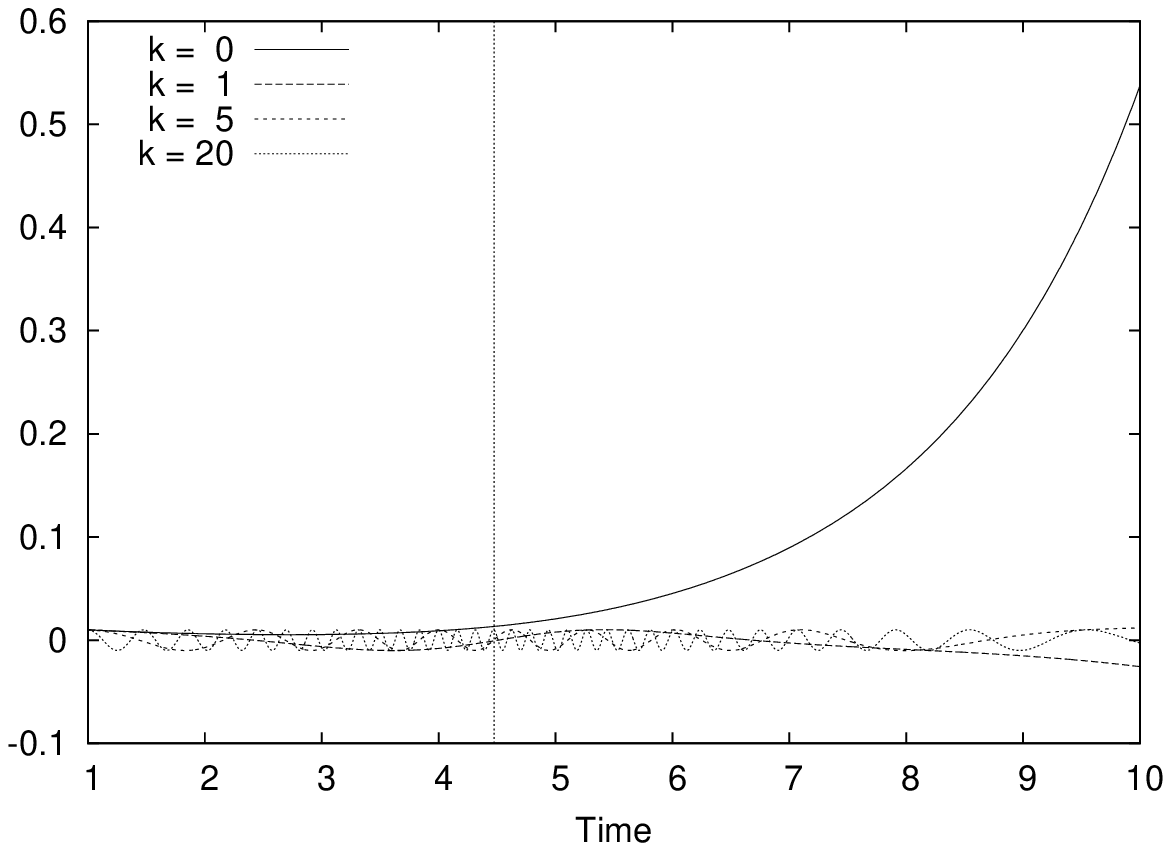} 
\caption{The growth of the density perturbations ${\cal D}$ in the background given by figure \ref{figFriedmann} for
the wave numbers $k=\tilde k/a_{0}=0,1,5$ and 20. 
Initially, at $t_0=1$, ${\cal D}=0.01$.
The time of the bounce is indicated with a dotted
vertical line.
}
\label{figPertFriedmann}
\end{figure}

The growth of the density perturbations for the radiation background in picture \ref{figFriedmann} is shown for different
wave-numbers in picture \ref{figPertFriedmann}. Initially the amplitudes of the perturbations are roughly
constant with a small dip in the amplitude of the $k=0$ mode, that serves as an approximation for the
long wavelength limit, before the bounce. 

\begin{figure}
\hskip 1cm
\epsfxsize=4.5in
\epsffile{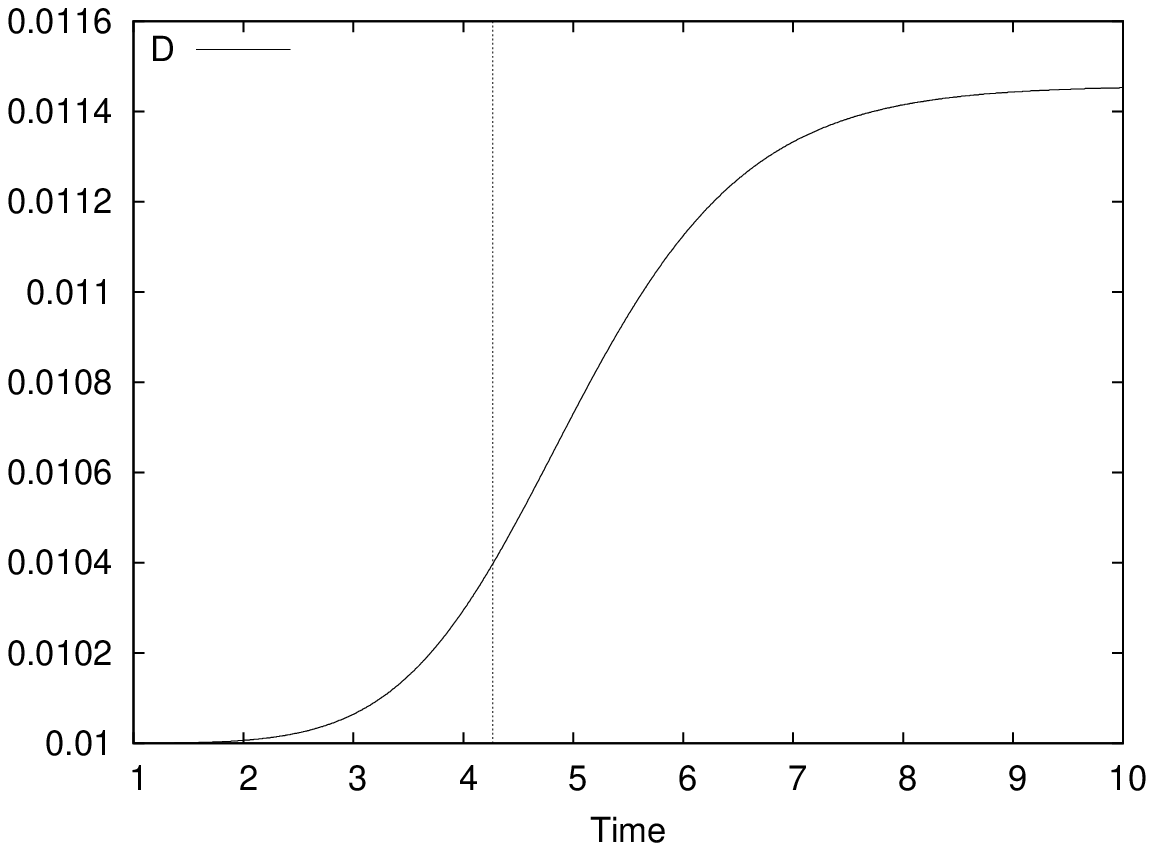} 
\caption{The growth of the density perturbations ${\cal D}$ for a dust background.
Initially, at $t_0=1$, ${\cal D}=0.01$. Initial conditions for background 
$\mu_0=0.001$, $\theta_0=-1.65$. $\Lambda=1$ and $C=9.35$.
The time of the bounce is indicated with a dotted
vertical line.
}
\label{figPertFriedmanndust}
\end{figure}

In the corresponding dust case the behaviour of the background quantities are similar to the radiation case, whereas
the evolution of the perturbations is independent of the wave-number, as can be seen from (\ref{eqDFriedmann}). See
\ref{figPertFriedmanndust} for the growth of  ${\cal D}$ . 

Bouncing closed Friedmann universes containing dust and radiation with cosmological constant are
excluded by observations, due to an argument by B\"orner and Ehlers \cite{Ehlers}. 
We here repeat their argument for the dust case in terms of our variables.
Using the
definition of the scale factor $a$ in terms of the expansion, $\theta\equiv3\dot a/a$, and Eq. (\ref{eqclosedmu})
one obtains 
$\mu=\mu_0(a_0/a)^3$, where the subscript 0 refers to present values. The constant in 
(\ref{eqclosedintegral})
is given by $3C\mu_0^{2/3}=3\mu_0+3\Lambda-\theta_0^2$ in terms of present values. At the
bounce we have $\dot a_{\star}=0$ and $\ddot a_{\star}>0$. Using this into Eqs. 
(\ref{eqclosedtheta})
and (\ref{eqclosedintegral}) one then obtains the following inequality 
\begin{equation}
\Omega_m\left[(z+1)^3-3(z+1)+2\right]<2
\end{equation}
where $\Omega_m\equiv 3\mu_0/\theta_0^2$ is the present fraction of dust relative to the
critical density and $z=a_0/a_{\star}-1$ is the redshift corresponding to the time of the bounce.
If the bounce took place before the formation
of quasars, that are observed at redshifts $z>4$, 
$\Omega _{m}$ cannot overcome the value $0.02$, in contradiction with observations.

%%%%%%%%%%%%%%%%%%%%%%%%%%%%%%%%%%%%%%%%%%%%%%%%

\end{document}